\documentclass[notitlepage,superscriptaddress,nofootinbib,tightenlines,preprintnumbers]{revtex4-1}
\usepackage{amsmath,verbatim,latexsym,amssymb,indentfirst,mathrsfs,mathtools,amsthm,bbm,bm,hyperref,url,cancel,subcaption,enumitem}
\usepackage[font=small,labelfont=bf,format=plain,justification=raggedright,singlelinecheck=false]{caption}
\usepackage{verbatim,indentfirst}
\usepackage[title,titletoc]{appendix}

\usepackage{graphicx}
\usepackage{epstopdf}
\newcommand{\caphead}[1]{{\bf #1}}

\renewcommand{\thesection}{\Roman{section}}
\renewcommand{\thesubsection}{\Roman{section} \Alph{subsection}}
\renewcommand{\thesubsubsection}{\Roman{section} \Alph{subsection} \arabic{subsubsection}}
\makeatletter
\def\p@subsection{}
\makeatother
\makeatletter
\def\p@subsubsection{}
\makeatother

\makeatletter
\newcommand\footnoteref[1]{\protected@xdef\@thefnmark{\ref{#1}}\@footnotemark}
\makeatother

\usepackage{soul}

\usepackage{color}

\newcommand{\KL}{{\rm KL}}

\def\id{\mathbbm{1}}   
\newcommand{\kB}{k_\mathrm{B}}  

\newcommand{\Sites}{N}  

\newcommand{\LParen}{ \bm{(} }
\newcommand{\RParen}{ \bm{)} }

\renewcommand\th{ {\rm th} }

\usepackage{nameref}
\usepackage{zref-xr}
\zxrsetup{toltxlabel}
\zexternaldocument*{SI_Learn}

\begin{document}
\title{Learning about learning about many-body physics}
%
\author{Weishun Zhong}
\email{wszhong@mit.edu. The first two coauthors contributed equally.}
\affiliation{Physics of Living Systems, Department of Physics, Massachusetts Institute of Technology, 400 Tech Square, Cambridge, MA 02139, USA}
\author{Jacob M. Gold}
\email{jacobmg@mit.edu}
\affiliation{Department of Mathematics, Massachusetts Institute of Technology, Cambridge, Massachusetts 02139, USA}
\author{Sarah Marzen}
\email{smarzen@kecksci.claremont.edu}
\affiliation{Physics of Living Systems, Department of Physics, Massachusetts Institute of Technology, 400 Tech Square, Cambridge, MA 02139, USA}
\affiliation{W. M. Keck Science Department,
Pitzer, Scripps, and Claremont McKenna Colleges,
Claremont, CA 91711, USA}
\author{Jeremy L. England}
\email{j@englandlab.com}
\affiliation{Physics of Living Systems, Department of Physics, Massachusetts Institute of Technology, 400 Tech Square, Cambridge, MA 02139, USA}
\affiliation{GlaxoSmithKline AI/ML, 200 Cambridgepark Drive, Cambridge MA, 02140, USA}
\author{Nicole Yunger Halpern}
\email{nicoleyh@umd.edu}
\affiliation{ITAMP, Harvard-Smithsonian Center for Astrophysics, Cambridge, MA 02138, USA}
\affiliation{Department of Physics, Harvard University, Cambridge, MA 02138, USA}
\affiliation{Research Laboratory of Electronics, Massachusetts Institute of Technology, Cambridge, Massachusetts 02139, USA}
\affiliation{Center for Theoretical Physics, Massachusetts Institute of Technology, Cambridge, Massachusetts 02139, USA}
\affiliation{Joint Center for Quantum Information and Computer Science, NIST and University of Maryland, College Park, MD 20742, USA}
\affiliation{Institute for Physical Science and Technology, University of Maryland, College Park, MD 20742, USA}
\date{\today}

%
%
\begin{abstract}
Diverse many-body systems, from soap bubbles to suspensions to polymers, learn and remember patterns in the drives that push them far from equilibrium. This learning may be leveraged for computation, memory, and engineering. Until now, many-body learning has been detected with thermodynamic properties, such as work absorption and strain. We progress beyond these macroscopic properties first defined for equilibrium contexts: We quantify statistical mechanical learning using representation learning, a machine-learning model in which information squeezes through a bottleneck. By calculating properties of the bottleneck, we measure four facets of many-body systems' learning: classification ability, memory capacity, discrimination ability, and novelty detection. Numerical simulations of a classical spin glass illustrate our technique. This toolkit exposes self-organization that eludes detection by thermodynamic measures: Our toolkit more reliably and more precisely detects and quantifies learning by matter while providing a unifying framework for many-body learning. \quad MIT-CTP/5297
\end{abstract}

{\let\newpage\relax\maketitle}

%
%
%
%

Many-body systems can learn and remember patterns of drives
that propel them far from equilibrium.
Such behaviors have been predicted and observed in many settings,
from charge-density waves~\cite{Coppersmith_97_Self,Povinelli_99_Noise}
to non-Brownian suspensions~\cite{Keim_11_Generic,Keim_13_Multiple,Paulsen_14_Multiple}, 
polymer networks~\cite{Majumdar_18_Mechanical}, 
soap-bubble rafts~\cite{Mukherji_19_Strength},
and macromolecules~\cite{Zhong_17_Associative}.
Such learning holds promise for engineering materials
capable of memory and computation.
Detecting such learning can also help us understand granular systems,
e.g., infer the history of forces experienced by an asteroid core.
This potential for applications, with experimental accessibility and ubiquity,
have earned these classical nonequilibrium many-body systems much attention recently~\cite{Keim_19_Memory}.

A classical, randomly interacting spin glass
exemplifies driven matter that learns.
Let us call a set $\{ \vec{A}, \vec{B}, \vec{C} \}$
of magnetic fields a \emph{drive}.
Consider randomly selecting a field from the drive
and applying it to the spin glass,
then repeating this process many times.
The spins absorb work from the fields.
The power absorbed shrinks adaptively, in a certain parameter regime:
The spins migrate toward a corner of configuration space 
where their configuration approximately withstands the drive's insults.
If new fields are imposed, the absorbed power spikes.
If fields from $\{ \vec{A}, \vec{B}, \vec{C} \}$ are reimposed,
the absorbed power spikes again, 
but less than under the unfamiliar fields~\cite{Gold_19_Self}.
The spin glass recognizes the original drive.

A simple, low-dimensional property of the material---absorbed power---distinguishes
drive inputs that fit a pattern from drive inputs that do not.
This property reflects a structural change in the spin glass's configuration.
The change is long-lived and not easily erased by new stimuli.
For these reasons, we say that the material has \emph{learned} the drive.

Many-body learning has been quantified with 
properties commonplace in thermodynamics.
Examples include power, as explained above,
and strain in polymers that learn stress amplitudes.
Such thermodynamic diagnoses offer insights
but suffer from two shortcomings.
First, the thermodynamic properties vary from system to system.
For example, work absorption characterizes the spin glass's learning;
strain characterizes non-Brownian suspensions'.
A more general approach would facilitate comparisons and standardize analyses.
Second, thermodynamic properties were defined for macroscopic equilibrium states.
Such properties do not necessarily describe
far-from-equilibrium systems' learning optimally.

\begin{figure}[hbt]
\centering
\includegraphics[width=.25\textwidth, clip=true]{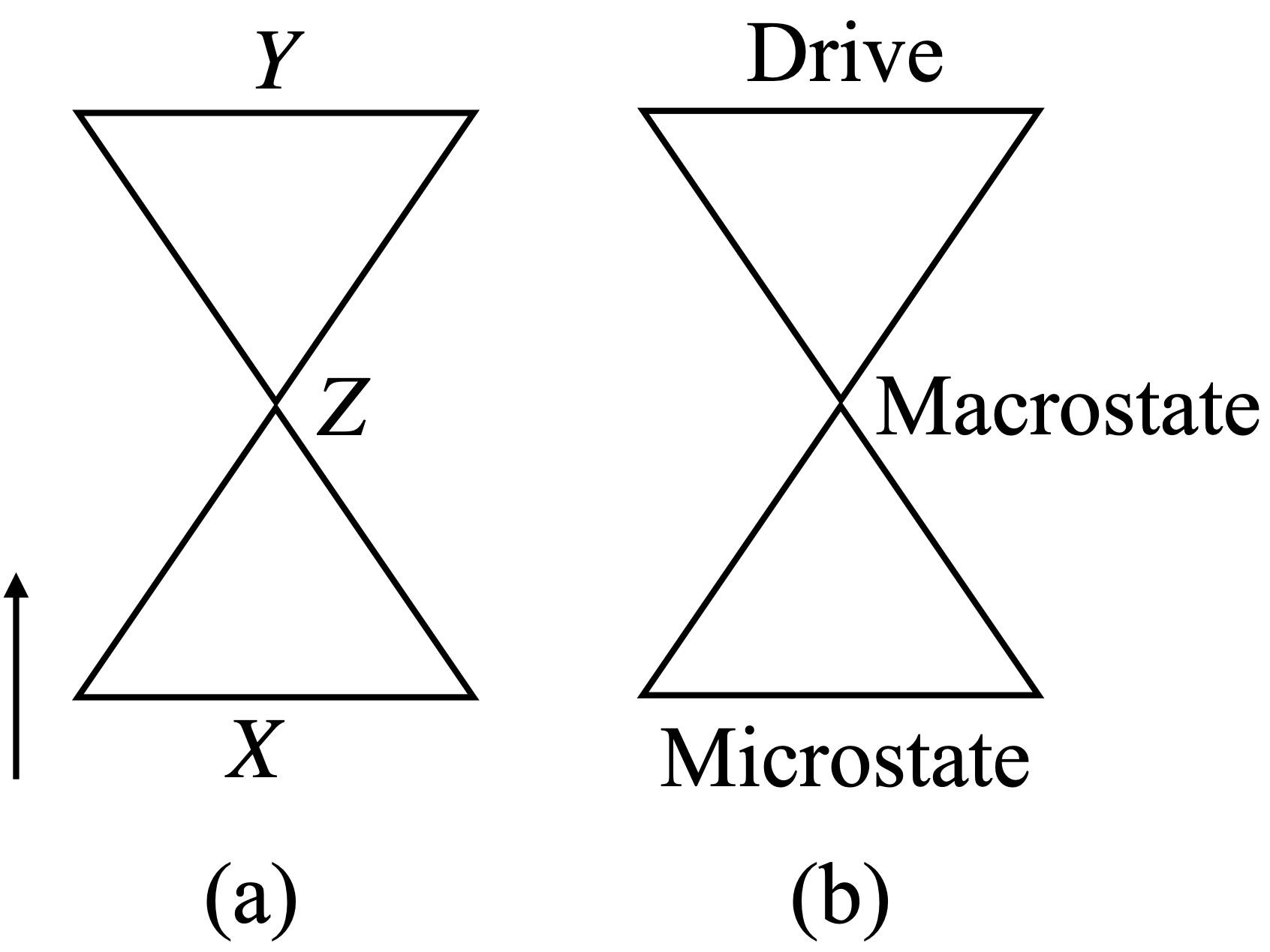}
\caption{\caphead{Parallel between two structures:}
(a) Bottleneck neural network, which performs representation learning.
(b) Nonequilibrium-statistical-mechanics problem.}
\label{fig_VAE_SM_Parallel}
\end{figure}

Separately from many-body systems' learning,
machine learning has flourished over the past decade~\cite{Nielsen_15_Neural,Goodfellow_16_Deep}.
Machine learning has helped elucidate how natural and artificial systems learn.
Neural networks developed over the past decade can undergo 
\emph{representation learning}~\cite{Bengio_12_Representation}
[Fig.~\ref{fig_VAE_SM_Parallel}(a)].
Such a neural network receives a high-dimensional variable $X$.
Examples include a sentence missing a word, e.g.,
``The \underline{\hspace{.5cm}} is shining.''
The neural network compresses the input
into a low-dimensional \emph{latent variable} $Z$,
e.g., word types and relationships.
The neural network decompresses $Z$ into a prediction $\hat{Y}$
of a high-dimensional variable $Y$.
In the example, $Y$ can be the word missing from the sentence,
and $\hat{Y}$ can be ``sun.''
The size of the bottleneck $Z$ controls a tradeoff
between the memory consumed and the prediction's accuracy.
We call the neural networks that perform representation learning
\emph{bottleneck neural networks}.

\begin{figure}[hbt]
\centering
\includegraphics[width=.95\textwidth, clip=true]{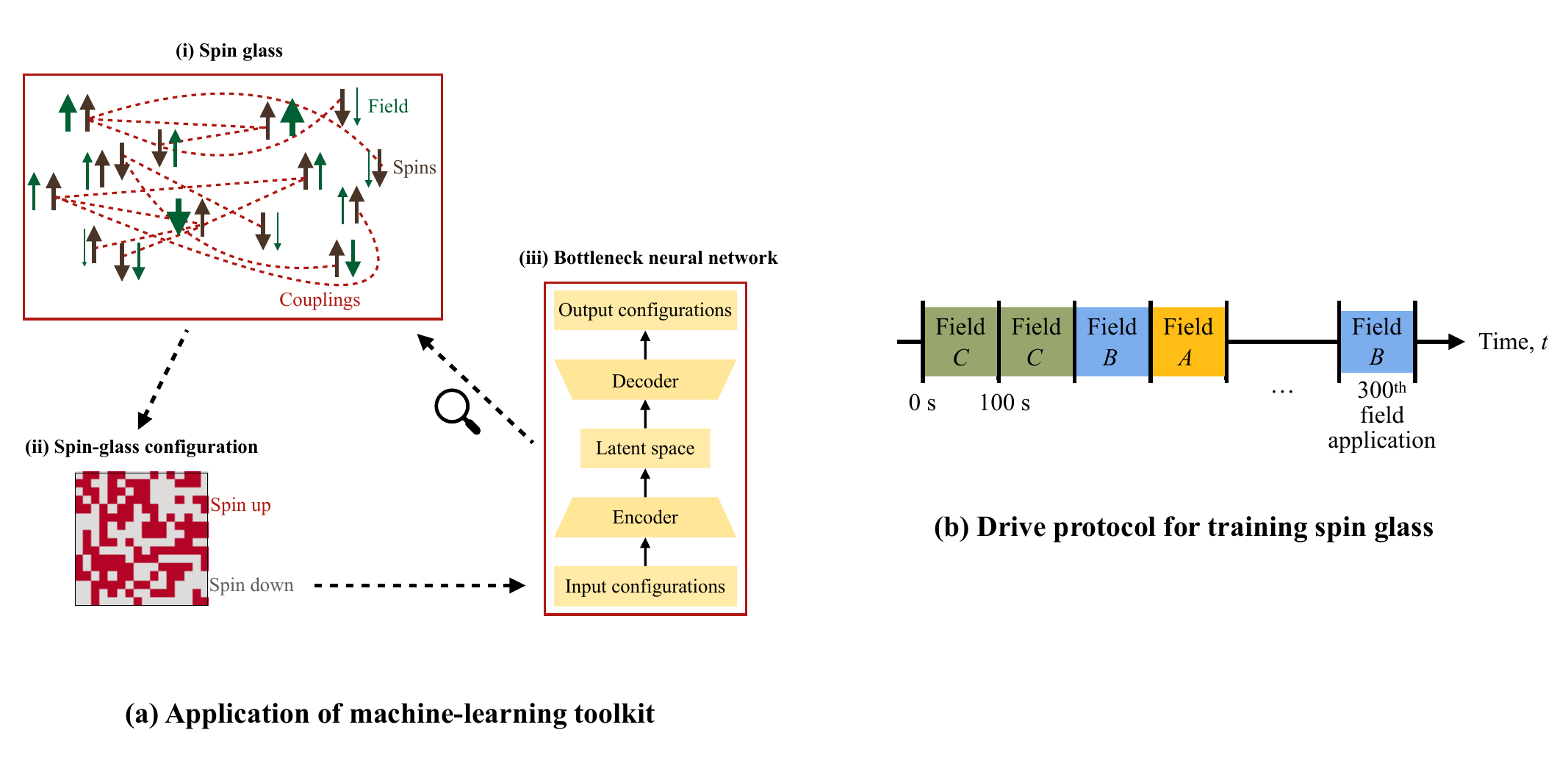}
\caption{\caphead{Schematic:}
Figure~\ref{fig_Schematics_Compiled}(a) sketches how
the many-body system is driven into configurations
on which the neural network trains.
Analyzing the neural network's latent space elucidates 
how much the many-body system has learned about its drive.
Figure~\ref{fig_Schematics_Compiled}(b) illustrates the protocol 
used to train the many-body system on a drive $\{A, B, C\}$.}
\label{fig_Schematics_Compiled}
\end{figure}

In this paper, we construct and deploy a bottleneck neural network 
to quantify how much many-body systems learn about 
the patterns of drives that force them:
We use representation learning to learn how much
many-body systems learn.
Our measurement protocols share the following structure
[Fig.~\ref{fig_Schematics_Compiled}(a)]:
The many-body system is trained with 
a drive (e.g., fields $\vec{A}$, $\vec{B}$, and $\vec{C}$).
Then, the system is tested (e.g., with a field $\vec{D}$).
Training and testing are repeated in many trials.
Configurations realized by the many-body system 
are used to train a bottleneck neural network via unsupervised learning.
Finally, we calculate properties of the neural network's bottleneck.
We illustrate with numerical simulations of the spin glass, 
whose learning has been detected with work absorption~\cite{Gold_19_Self}.
Our methods generalize to other platforms, however.
This machine-learning toolkit offers three advantages:
\begin{enumerate} 
   \item 
   Bottleneck neural networks register learning behaviors 
   more thoroughly and precisely
   than work absorption.

   \item 
   Our framework encompasses a wide class of
   strongly driven many-body systems.
   Although we illustrate with the example of a spin glass,
   the framework does not rely on any particular thermodynamic property
   tailored to spins. Our neural network scores a many-body system's learning behaviors with dimensionless numbers that can be compared across platforms.

   \item 
   Our approach unites machine learning with 
   learning by many-body systems.
   The union is conceptually satisfying.
   
\end{enumerate}
We measure four facets of many-body learning:
classification ability, memory capacity, discrimination ability, and novelty detection.
Our techniques, however, can be extended to other facets.

%
%
%
\section{Results}

First, we introduce our bottleneck neural network.
Then, we define the spin glass on which we will test 
our machine-learning toolkit.
We finally show how to quantify, using representation learning, 
how much a many-body system learns about a drive.

\subsection*{Bottleneck neural network}


Representation learning, we argue, shares its structure with
a problem in nonequilibrium statistical mechanics
[Fig.~\ref{fig_VAE_SM_Parallel}(b)].
Consider a many-body system subject to a strong drive.
The system's microstate occupies a high-dimensional space, 
like the input $X$ to a bottleneck neural network.
A macrostate synopsizes the microstate in a few numbers, 
such as particle number and magnetization.
This synopsis parallels the latent space $Z$. 
If the many-body system has learned the drive, 
the macrostate encodes the drive.
One may reconstruct the drive from the macrostate,
as a bottleneck neural network reconstructs $Y$ from $Z$.\footnote{
See~\cite{Alemi_18_TherML} for a formal parallel 
between representation learning and equilibrium thermodynamics.}

We construct a neural network inspired by this parallel.
As the macrostate informs computations in the statistical-mechanics problem described above, the neural network's bottleneck informs our computations.
One might initially aim for a bottleneck neural network 
that predicts drives from configurations $X$.
But such a neural network would undergo supervised learning,
if constructed according to the state of the art of when this paper was written.
During supervised learning, the neural network receives tuples 
(configuration of the many-body system, label of drive that generated the configuration).
The drive labels are not directly available to the many-body system.
So successful predictions by neural network predictions would not necessarily reflect 
only learning by the many-body system. 
Hence we design a bottleneck neural network that performs unsupervised learning,
receiving only configurations.

This neural network is a \emph{variational autoencoder},~\cite{Kingma_13_Auto,JR_14_Stochastic,Doersch_16_Tutorial},
a generative model:
It receives samples $x$ from a distribution over
the possible $X$ values,
creates a variational model for the distribution, 
and samples from the model.
The model is refined via Bayesian variational inference
(see App.~\ref{sec_NN_Details} for an overview).
The model's parameters are optimized via backpropagation during training.

Our variational autoencoder has five fully connected hidden layers,
with neuron numbers 200-200-(number of $Z$ neurons)-200-200.
We usually restrict the latent variable $Z$ to 2-4 neurons.
This choice enables us to visualize the latent space
and suffices to quantify the spin glass's learning.
Growing the many-body system
may require more latent dimensions,
as may growing the number of drives whose patterns 
the many-body system must learn.
But our studies suggest that the number of dimensions needed
$\ll$ the system size.

\begin{figure}[hbt]
\centering
\includegraphics[width=.5\textwidth, clip=true]{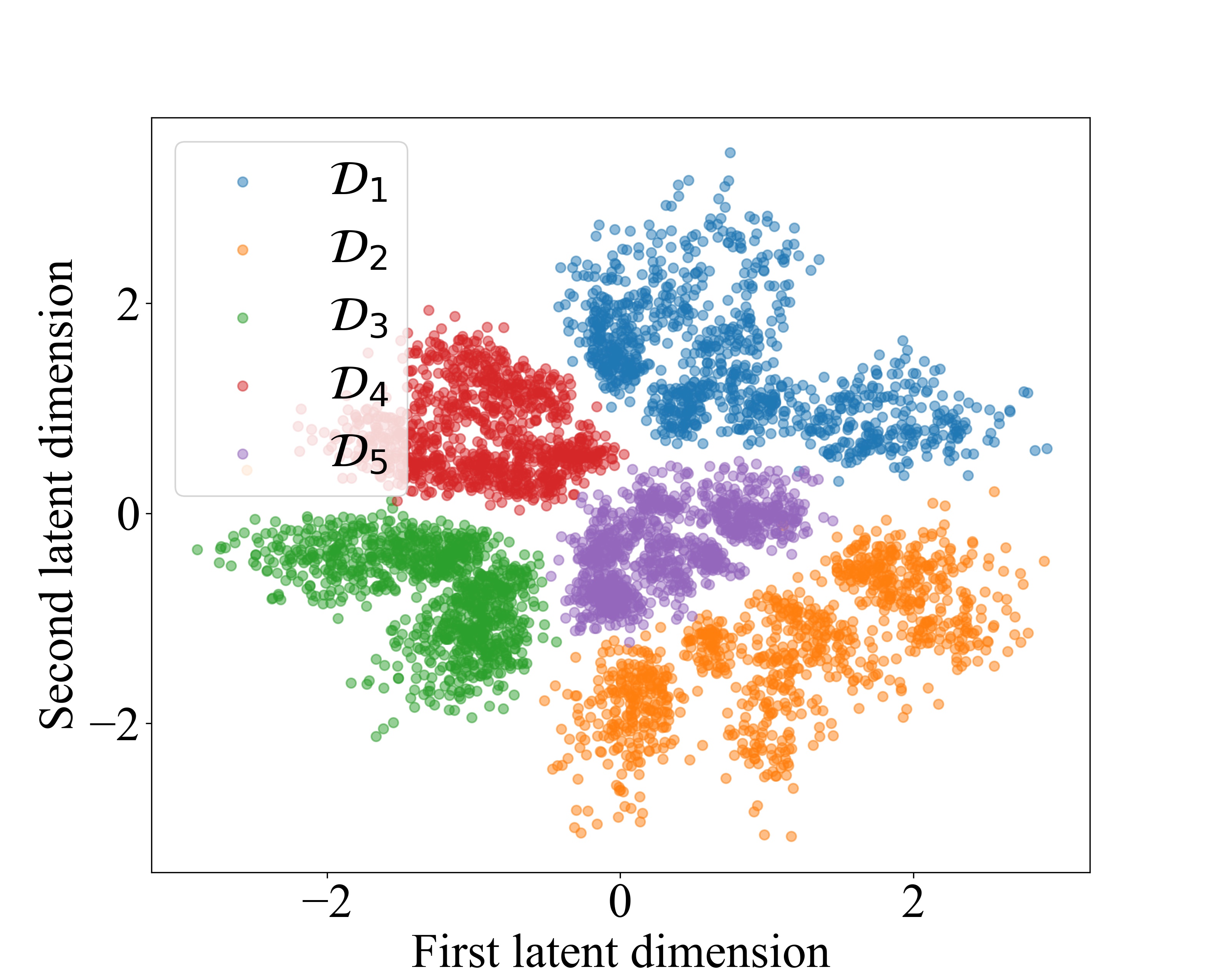}
\caption{\caphead{Visualization of latent space, $Z$:}
$Z$ consists of neurons $Z_1$, represented along the $x$-axis, 
and $Z_2$, represented along the $y$-axis.
A variational autoencoder formed $Z$
while training on configurations assumed by a 256-spin glass
exposed to different drives in different trials.
The neural network mapped each configuration to a dot in latent space.
After the training completed, each dot was colored according to 
which field was applied when the configuration was realized.
}
\label{fig_Latent_Space}
\end{figure}

Figure~\ref{fig_Latent_Space} depicts the latent space $Z$.
Each neuron corresponds to one axis
and represents a continuous-valued real number.
The latent space was formed via the protocol detailed below, in the section
``How to quantify a many-body system's learning of a drive,
using representation learning.''
To synopsize, we trained the spin glass on one drive in each of 1,000 trials;
trained the spin glass in another drive in each of 1,000 trials;
and so on, for five drives total.
On the end-of-trial spin-glass configurations, the neural network was trained.
The neural network compressed each configuration to a dot in latent space.
We colored each dot according to which drive produced 
the corresponding configuration.
We added the colors after the neural network's training, 
so the neural network received no configurations' drive labels.
Same-color dots cluster together,
so the spin glass distinguished the drives,
as recognized by the neural network.


%
%

One might wonder whether our toolkit requires deep learning.
Could simpler algorithms detect and measure many-body learning
as sensitively?
Appendix~\ref{sec_Justify_ML} responds negatively.
We compare our neural network with simpler competitors
that perform unsupervised learning:
a single-layer linear neural network,
related to principal-component analysis~\cite{Bourland_88_Auto},
and a clustering algorithm.
The bottleneck neural network outperforms both competitors.
(Competitors that perform supervised learning would enjoy
an unfair advantage and, as explained above,
would not reflect the many-body's system learning faithfully.)

\subsection*{Spin glass}


A spin glass exemplifies the many-body learner~\cite{Gold_19_Self}.
We illustrate our machine-learning toolkit
by simulating a glass of $\Sites = 256$ classical spins.
The $j^\th$ spin occupies one of two possible states:
$s_j = \pm 1$.

The spins couple together and experience an external magnetic field:
Spin $j$ evolves under a Hamiltonian
\begin{align}
   \label{eq_Hamiltonian_j}
   H_j(t)
   =  \sum_{k \neq j}  J_{jk}  s_j  s_k
   +  A_j(t) s_j ,
\end{align}
and the spin glass evolves under
$H(t)  =  \frac{1}{2} \sum_{j = 1}^\Sites  H_j(t)$,
at time $t$.
We call the first term in Eq.~\eqref{eq_Hamiltonian_j} the \emph{interaction energy} 
and the second term the \emph{field energy}.
The couplings $J_{j k}  =  J_{kj}$ are defined through
an Erd\"{o}s-R\'enyi random network:
Spins $j$ and $k$ have some probability $p$ 
of interacting, for all $j$ and $k \neq j$.
Each spin couples to eight other spins, on average.
The nonzero couplings $J_{j k}$ are selected according to 
a normal distribution of standard deviation 1.

$A_j(t)$ denotes the magnitude and sign
of the external field experienced by spin $j$ at time $t$.
The field always points along the same direction (the $z$-axis),
so we omit the arrow from $\vec{A}_j(t)$.
We will simplify the notation for the field from $\{ A_j(t) \}_j$ to $A$
(or $B$, etc.).
Each $A_j(t)$ is selected according to 
a normal distribution of standard deviation 3.
The field changes every 100 seconds. 

To train the spin glass, we construct a drive
by constructing a set $\{A, B, \ldots \}$ of random fields.
We randomly select a field from the set, then apply the field for 100 s.
This selection-and-application process is performed 300 times
[Fig.~\ref{fig_Schematics_Compiled}(b)]. 

The spin glass exchanges heat with
a bath at a temperature $T = 1 / \beta$.
We set Boltzmann's constant to $\kB = 1$.
Energies are measured in Kelvins (K).
To flip, a spin must overcome a height-$B$ energy barrier.
Spin $j$ tends to flip at a rate 
$\omega_j 
   =  e^{\beta [ H_j(t) - B]}
   / (1 \text{ second})  \, .$
This rate has the form of Arrhenius's law
and obeys detailed balance.
The average spin flips once per $10^7$ s.
We model the evolution with discrete 100-s time intervals,
using the Gillespie algorithm.

The spins absorb work when the field changes,
as from $\{ A_j(t) \}$ to $\{ A'_j(t) \}$. 
The change in the spin glass's energy equals
the work absorbed by the spin glass:
$W  :=  \sum_{j = 1}^\Sites  
   \left[  A'_j(t)  -  A_j(t)  \right]  s_j .$
Absorbed power is defined as $W / ( \text{100 s} )$. 
The spin glass dissipates heat by
losing energy as spins flip.

The spin glass is initialized in a uniformly random configuration 
$\mathcal{C}$.
Then, the spins relax in the absence of any field for 100,000 seconds.
The spin glass navigates to near a local energy minimum.
If a protocol is repeated in multiple trials, 
all the trials begin with the same configuration $\mathcal{C}$.

In a certain parameter regime, the spin glass learns its drive effectively,
even according to the absorbed power~\cite{Gold_19_Self}.
Consider training the spin glass on a drive $\{ A, B, C \}$.
The spin glass absorbs much work initially.
If the spin glass learns the drive, the absorbed power declines.
If a dissimilar field $D$ is then applied, the absorbed power spikes.
If the familiar fields are reapplied, the absorbed power spikes again,
but less.
The spin glass learns effectively in the ``Goldilocks regime'' of
$\beta = 3$ K$^{-1}$ and $B = 4.5$ K~\cite{Gold_19_Self}:
The temperature is high enough,
and the barriers are low enough,
that the spin glass can explore phase space.
But $T$ is low enough, and the barriers are high enough,
that the spin glass is not hopelessly peripatetic.

Spins can fail to learn nontrivially,
yet adopt configurations that reflect a drive.
For example, the spins can be entrained to the field.
The spins would bear the field's stamp as silly putty bears a thumbprint.
A thumbprint vanishes as soon as the silly putty is smoothed.
Hence the silly putty undergoes no long-lived structural change 
that resists erasure;
the silly putty does not learn robustly.
Alternatively, most of the spins can remain frozen, while only a few flip.
One might infer the drive from the few flippable spins,
though most of the glass would contain no information about the drive.
We confirm that our spin glass does not exhibit these behaviors,
in App.~\ref{sec_Not_Enslaved_Or_Frozen}:
The spin glass's learning is nontrivial.

\subsection*{How to quantify a many-body system's learning of a drive,
using representation learning}

We detect and quantify four facets of learning: 
classification ability, memory capacity, discrimination, and novelty detection.
One \emph{classifies} a stimulus by answering the question
``Which of the possible stimuli is this one?''
A system's \emph{memory capacity} is the number of fields
that the system can remember.\footnote{
(We use the term ``memory capacity'' in the physical sense of~\cite{Keim_19_Memory}.
A more specific, technical definition of ``memory capacity''
is used in reservoir computing~\cite{Jaeger_02_Short}.)}
One performs \emph{novelty detection} by answering the question
``Have I encountered this stimulus before?''
One \emph{discriminates} between stimuli $A$ and $B$ by answering
``How much of the present stimulus consists of $A$, 
and how much consists of $B$?''

Below, we illustrate the application of our toolkit
by quantifying classification ability.
The Methods show how to apply our toolkit to the other three facets of learning.
Further facets may be quantified similarly.
Our machine-learning approach detects and measures learning
more reliably and precisely than absorbed power does.
Code used and data generated are accessible at~\cite{Github_repo}.




A system \emph{classifies} a stimulus when 
identifying the stimulus as one of many possibilities.
First, we detail the protocol run on the spin glass.
Second, we show how to measure
the spin glass's classification ability using representation learning.
Third, we measure the spin glass's classification ability
using absorbed power.
The neural network, we find, reflects more of 
the spin glass's classification ability
than absorbed power does.

The spin glass underwent the following protocol.
We generated random fields $A$, $B$, $C$, $D$, and $E$.
From 4 of the fields, we formed the drive $\mathcal{D}_1  :=  \{A, B, C, D\}$.
On the drive, we trained the spin glass in each of 1,000 trials.
In each of 1,000 other trials, we trained a refreshed spin glass on
a drive $\mathcal{D}_2  :=  \{A, B, C, E\}$.
We repeated this process for each of the 5 possible 4-field drives.
Ninety percent of the trials were randomly selected 
for training the neural network.
The rest were used for testing.

We measured the spin glass's ability to classify drives,
using the neural network, as follows:
We fixed a time $t$, then identified the configurations 
occupied by the spin glass at $t$ in the spin-glass-training trials.
On these configurations, we trained the neural network.
The neural network populated the latent space with dots
(similarly to in Fig.~\ref{fig_Latent_Space}).
The dots generated by drive $\mathcal{D}_j$
approximated a probability density $P_j$,
for each of $j = 1, 2, 3, 4, 5$.

\begin{figure}[hbt]
\centering
\includegraphics[width=.5\textwidth, clip=true]{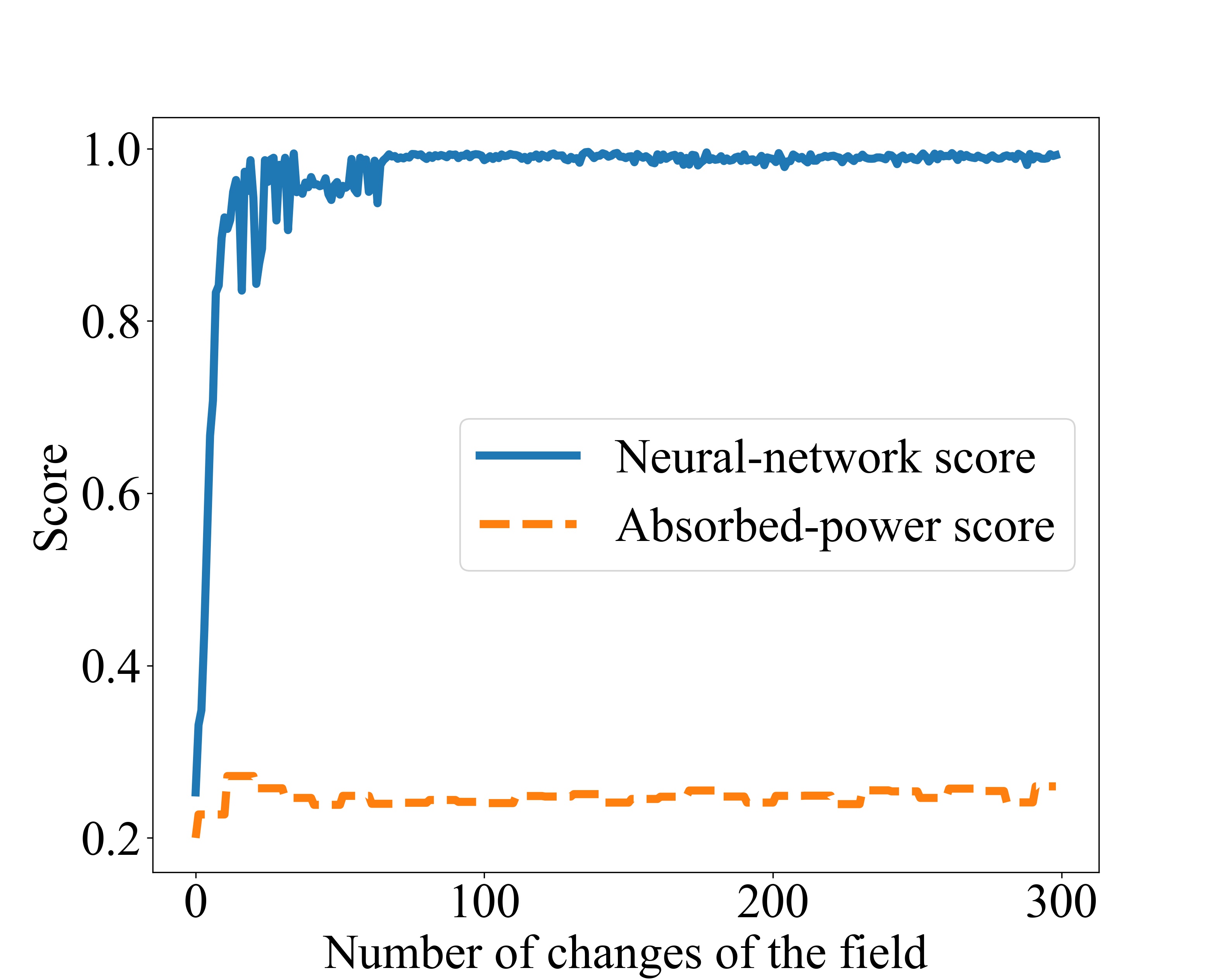}
\caption{\caphead{Quantification of a many-body system's classification ability:}
A spin glass classified a drive as one of five possibilities.
The blue, upper curve represents the system's classification ability,
as quantified by a bottleneck neural network.
The orange, lower curve represents the classification ability
as quantified with the absorbed power.
The neural-network score rises to near the maximum, 1.00.
The thermodynamic score exceeds the random-guessing score, $1/5$, slightly.
The neural network therefore detects more of the spins' classification ability.
}
\label{fig_Classification}
\end{figure}

We then gave the neural network
a time-$t$ configuration from a test trial.
The neural network compressed the configuration into a latent-space point.
We calculated the probability that drive $\mathcal{D}_j$ generated that point,
using $P_j$, for all $j$.
The highest-probability drive most likely generated the point,
by maximum-likelihood estimation~\cite{Bishop_06_Pattern}.
We performed this testing and estimation for each trial in the test data.
The fraction of trials in which the estimation  succeeded
constitutes the \emph{score}.
The score is plotted against $t$ in Fig.~\ref{fig_Classification}
(blue, upper curve).

We compare with the classification ability
attributed to the spin glass by the absorbed power:
We fixed a drive $\mathcal{D}_j$ and a time $t$.
We identified the neural-network-training trials in which
$\mathcal{D}_j$ was applied at time $t$.
From the power absorbed then, we formed a histogram.
We performed this process for each drive $\mathcal{D}_j$.
Then, we took a trial from the test set
and identified the power absorbed at $t$.
We inferred which drive most likely produced that power,
applying maximum-likelihood estimation to the histograms. 
The guess's score appears as the orange, lower curve
in Fig.~\ref{fig_Classification}.

A score maximizes at 1.00 if the drive is always guessed accurately.
The score is lower-bounded by the random-guessing value
$1 / (\text{number of drives}) = 1/5$.
In Fig.~\ref{fig_Classification}, each score grows 
over tens of field switches.
The absorbed-power score begins at\footnote{
\label{foot_Why_Not_0.2}
The neural network's score begins slightly above 0.20. 
One might expect the score to begin at 0.20:
At $t = 0$, the spin glass has not experienced the drive,
so the neural network receives no information about the drive,
so the neural network can guess the drive only randomly.
The distance from 0.20, we expect, comes from stochasticity of three types:
the spin glass's initial configuration, maximum-likelihood estimation,
and stochastic gradient descent. 
Stochasticity of only the first two types affects the absorbed-power score.}
0.20 and comes to fluctuate around 0.25.
The neural network's score comes to fluctuate slightly below 1.00.
(The neural network's score begins slightly above 0.20. 
One might expect the score to begin at 0.20:
At $t = 0$, the spin glass has not experienced the drive,
so the neural network receives no information about the drive,
so the neural network can guess the drive only randomly.
The distance from 0.20, we expect, comes from stochasticity of three types:
the spin glass's initial configuration, maximum-likelihood estimation,
and stochastic gradient descent. 
Stochasticity of only the first two types affects the absorbed-power score.)
Hence the neural network detects more of the spin glass's classification ability
than the absorbed power does,
in addition to suggesting a means of quantifying the classification ability rigorously.
Having illustrated our machine-learning toolkit with classification,
we detail applications to memory capacity, novelty detection,
and discrimination in the Methods.

\section{Discussion}

We have detected and quantified a many-body system's learning 
of drive patterns, using representation learning.
Our toolkit affords greater sensitivity than absorbed power,
a representative of the thermodynamic toolkit applied
to detect many-body learning until now.
Our technique quantifies classification ability,
memory capacity, discrimination ability, and novelty detection.
The toolkit is general, not relying on whether 
the system exhibits magnetization or strain or another thermodynamic response.
The Methods establish the feasibility of applying our toolkit
in a variety of experiments and simulations.
This approach provides a framework for understanding memory---a 
basic, widely realized, and usable trait---in
a unified manner across classical statistical mechanics.
This framework opens several opportunities for future research;
we detail two below.

First, our toolkit is well-suited to more open problems about many-body learners.
An example problem concerns the soap-bubble raft in~\cite{Mukherji_19_Strength}.
Experimentalists trained a raft of soap bubbles with
an amplitude-$\gamma_{\rm t}$ strain.
The soap bubbles' positions were tracked,
and variances in positions were calculated.
No such measures distinguished trained rafts
from untrained rafts;
only stressing the raft and reading out the strain could~\cite{Mukherji_19_Strength,Miller_19_Raft}.
Our bottleneck neural network is well-poised to identify 
microscopic properties that distinguish trained from untrained rafts.
Similarly, representation learning may facilitate 
the detection of active matter.
Self-organization is detected now through 
simple, large-scale, easily visible signals~\cite{Heylighen_02_Science}.
Bottleneck NNs could identify patterns
invisible in thermodynamic measures.

Second, in statistical mechanics, we parameterize macrostates with
volume, energy, magnetization, and other thermodynamic variables.
Macrostates in statistical mechanics parallel 
the latent space in our bottleneck neural network
(Fig.~\ref{fig_VAE_SM_Parallel}).
Which variables parameterize the neural network's latent space?
Latent space may suggest definitions of new thermodynamic variables,
or hidden relationships amongst known thermodynamic variables.

\begin{figure}[hbt]
\centering
\includegraphics[width=.95\textwidth, clip=true]{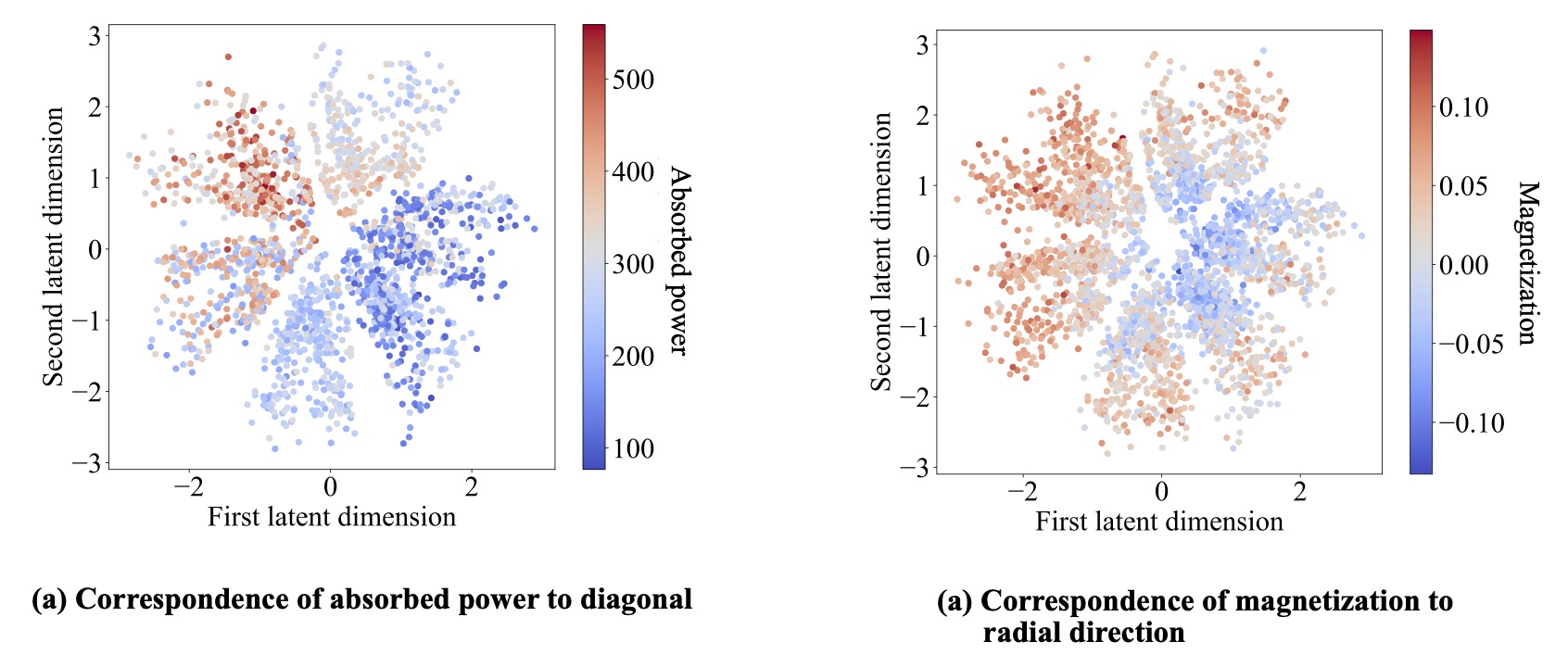}
\caption{\caphead{Correspondence of latent-space directions
to thermodynamic quantities:}
A variational autoencoder trained on the configurations assumed by a spin glass 
exposed to fields $A$, $B$, and $C$.
We have color-coded each latent-space plot, 
highlighting how a thermodynamic property changes 
along some direction.
In Fig.~\ref{fig_Decoding_Z_Compiled}(a), 
the absorbed power grows
from the bottom righthand corner to the upper lefthand corner.
In Fig.~\ref{fig_Decoding_Z_Compiled}(b), 
the magnetization grows radially.}
\label{fig_Decoding_Z_Compiled}
\end{figure}

We illustrate by training the spin glass with a drive $\{ A, B, C \}$ in each of many trials.
On the end-of-trial configurations, we trained the neural network.
Two latent-space directions have physical significances,
as shown in Fig.~\ref{fig_Decoding_Z_Compiled}:
The absorbed power grows along the diagonal from the bottom righthand corner
to the upper lefthand corner [Fig.~\ref{fig_Decoding_Z_Compiled}(a)].
The magnetization grows radially [Fig.~\ref{fig_Decoding_Z_Compiled}(b)].
The directions are nonorthogonal, suggesting 
a nonlinear relationship between the thermodynamic variables.
Convention biases physicists toward measuring
volume, magnetization, heat, work, etc.
The neural network may identify new macroscopic variables
better-suited to far-from-equilibrium statistical mechanics,
or nonlinear relationships amongst thermodynamic variables.

We can translate, as follows, between conventional thermodynamic variables and the latent-space directions $z_1$ and $z_2$: 
List the conventional thermodynamic variables expected to be relevant: $v_1, v_2, \ldots, v_n$. 
For example, $v_1$ may denote the work absorption, and $v_2$ may denote the magnetization. The neural network populates the latent space with dots during training. Each dot corresponds to $v_j$'s calculable from the corresponding many-body configuration. A feedforward neural network can decompose each $z_k$ as a function of the $v_j$'s. We will have decomposed the latent-space variables in terms of thermodynamic variables, translating between the two.
A bottleneck neural network could uncover new theoretical physics,
as discussed in, e.g.,~\cite{Carleo_19_Machine,Wu_19_Toward,Iten_20_Discovering}.

%
%
%
\section{Methods}

In the Results, we applied our machine-learning toolkit
to quantify classification ability.
Here, we apply the toolkit to quantify three more facets of learning:
memory capacity, discrimination, and novelty detection.
We also demonstrate the feasibility of applying our toolkit to experiments.

\subsection*{Memory capacity: How many fields can the system remember?}
\label{sec_Capacity}


How many fields can a many-body system remember?
A bottleneck neural network, we find, 
registers a greater memory capacity 
than absorbed power registers.
Hence the neural network reflects statistical mechanical learning,
at high field numbers,
that the absorbed power does not.

We illustrate by constructing 50 random fields.
We selected 40 to form a drive $\mathcal{D}_1$,
selected 40 to form a drive $\mathcal{D}_2$,
and repeated until forming 5 drives.
We trained the spin glass on drive $\mathcal{D}_j$
in each of 1,000 trials, for each of $j = 1, 2, \ldots 5$.
Ninety percent of the trials were designated as neural-network-training trials;
and 10\%, as neural-network-testing trials.

The choice of 50 fields is explained in 
App.~\ref{app_Capacity_Work}:
Fifty fields exceed the spin-glass capacity registered by the absorbed power.
We will show that 50 fields do not exceed the capacity
registered by the neural network:
The neural network identifies spin-glass learning missed by the absorbed power.

\begin{figure}[hbt]
\centering
\includegraphics[width=.5\textwidth, clip=true]{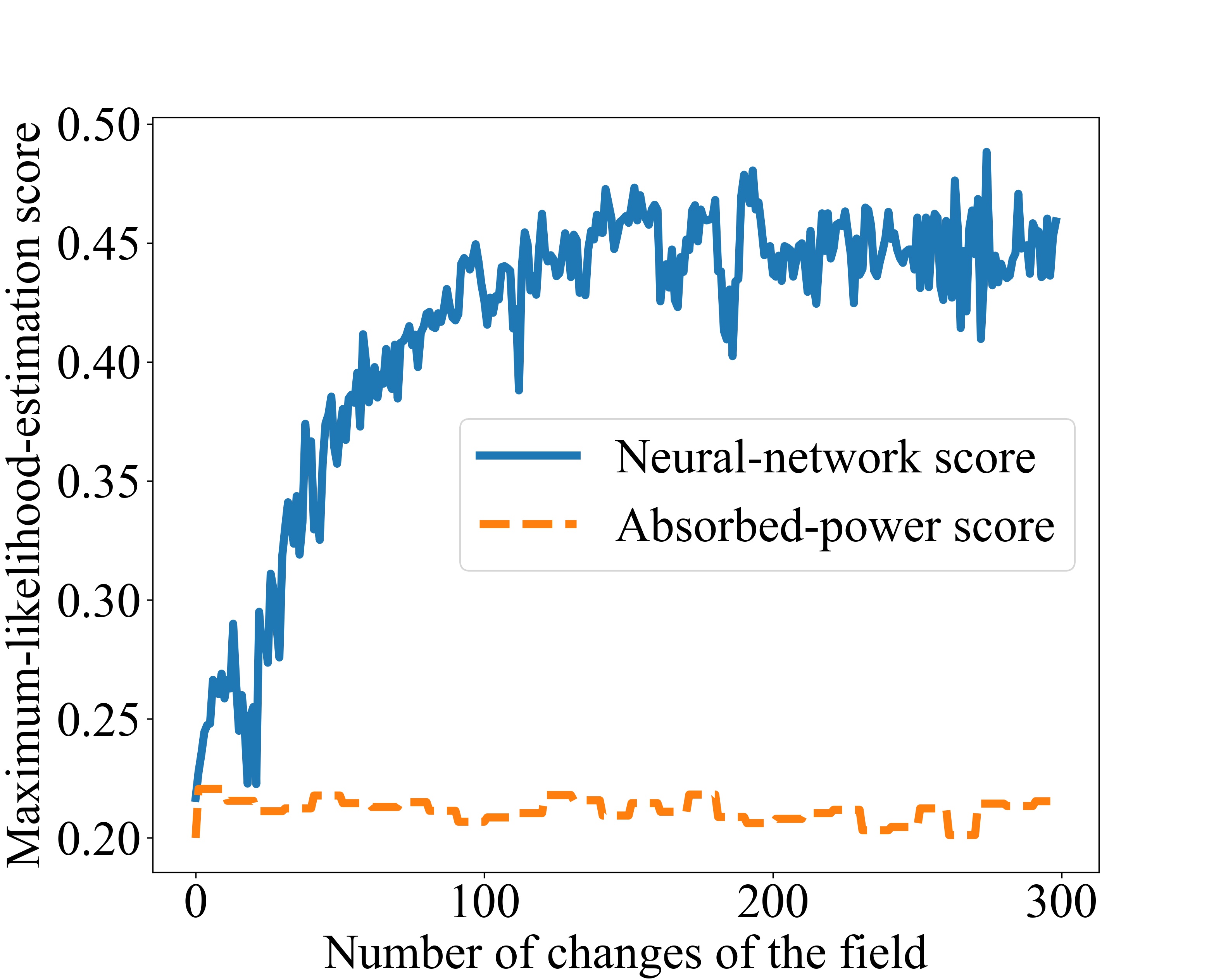}
\caption{\caphead{Quantification of memory capacity:}
A spin glass was trained on one of five drives in each of many trials.
Each drive was formed from 40 fields selected from 50 random fields.
The upper, blue line represents the memory capacity
attributed to the spin glass by a bottleneck neural network.
The lower, orange line represents the memory capacity
attributed by absorbed power.}
\label{fig_Capacity}
\end{figure}

We used representation learning to quantify the spin glass's capacity as follows.
For a fixed time $t$, we collected 
the configurations occupied by the spin glass at $t$
in the neural-network-training trials.
On these configurations, the neural network performed unsupervised learning.
The neural network populated its latent space with dots
that formed five clusters.
The cluster sourced by drive $\mathcal{D}_j$
approximated a probability density $P_j$.
We fed the neural network the configuration occupied at $t$ during a test trial.
The neural network formed a new dot in latent space.
We estimated the probability that drive $\mathcal{D}_j$ formed the drive,
using $P_j$, for each $j$.
The greatest probability stemmed from the drive $\mathcal{D}_j$
that most likely, according to the neural network, produced the point.
That is, we applied maximum-likelihood estimation.
The fraction of test trials in which the neural network guessed correctly
constitutes the neural network's score.
The score is plotted against $t$ in Fig.~\ref{fig_Capacity},
as the blue, upper curve.

The neural network's score is compared with the absorbed power's score,
calculated as follows.
For a fixed time $t$, we identified the power absorbed at $t$
in each neural-network-testing trial.
We histogrammed the power absorbed
when $\mathcal{D}_j$ was applied at $t$,
for each $j = 1, 2, \ldots, 5$.
We then identified the power absorbed at $t$ in a test trial.
Comparing with the histograms, we inferred which drive
was most likely being applied.
We repeated this inference with each other test trial.
In which fraction of the trials did the absorbed power identify the drive correctly?
This number forms the absorbed power's score.
The score is plotted as the lower, orange curve in Fig.~\ref{fig_Capacity}.

The higher the score, the greater the memory capacity
attributed to the spin glass.
The absorbed power identifies the drive 
in approximately $20\%$ of the trials,
as would random guessing.
The score remains approximately constant,
because the number of fields exceeds the spin-glass capacity
measured by the absorbed power.
In contrast, the neural network's score grows over
$\approx 150$ changes of the field,
then plateaus at $\approx 0.450$.
The neural network points to the wrong drive most of the time
but succeeds significantly more often than the absorbed power.
Hence representation learning uncovers
more of the spin glass's memory capacity
than absorbed power measure does.

In summary, a many-body system's memory capacity can be quantified as
the greatest number of fields in any drive 
on which maximum-likelihood estimation, 
based on a neural network's latent space,
scores better than random guessing.


%
%
%
\subsection*{Discrimination: How new is this field?}
\label{sec_Discriminate}


Suppose that a many-body system learns fields $A$ and $B$,
then encounters a field that interpolates between them.
Can the system recognize that 
the new field contains familiar constituents?
Can the system discern how much $A$ contributes
and how much $B$ contributes?
The answers characterize the system's discrimination ability,
which we quantify with a maximum-likelihood-estimation score.
Estimates formed from the neural network's latent space
reflect more of the system's discriminatory ability
than do estimates formed from absorbed power.

We illustrate with the spin glass, forming a drive $\{A, B, C\}$.
Each trial began with 300 subsequent time intervals.
In each interval, a field was selected randomly from the drive and applied.
The spin glass was then tested with
a linear combination $D_w = w A + (1 - w) B$.
The weight $w$ varied from 0 to 1, in steps of $1/6$, across trials.

We measured the spin glass's discrimination using the neural network as follows.
We identified the final configuration assumed 
by the spin glass in each trial.
These configurations were split into neural-network-training data
and neural-network-testing data.
The training trials ended with configurations
on which the neural network was trained.
Then, the neural network received a configuration with which
a neural-network-testing trial ended.
The neural network mapped the configuration to a latent-space point.
We inferred which field most likely generated that point,
using maximum-likelihood estimation on the latent space.
We tested the neural network with all the test trials. 
The fraction of maximum-likelihood estimates that were correct
formed the neural network's score.

Similarly, we measured the spin glass's discrimination
using the absorbed power.
We fixed a value of $w$, then identified the neural-network-training trials
that ended with the application of $D_w$.
We identified the power $\mathcal{P}$ absorbed by the spin glass
after the $D_w$ application.
We histogrammed $\mathcal{P}$, inferring the probability that,
if shown $D_w$ for a given $w$,
the spin glass will absorb an amount $\mathcal{P}$ of power.
We formed a histogram for each value of $w$.
Then, we calculated the power absorbed
during a neural-network-testing trial.
We inferred which field most likely generated that point,
applying maximum-likelihood estimation to the histograms.
We repeated the maximum-likelihood estimation with 
each neural-network-testing trial.
The absorbed power's score equals
the fraction of the trials in which 
the maximum-likelihood estimation was correct.

The neural network's score equals about double the absorbed power's score,
for latent spaces of dimensionality 2 to 20.
The neural network scores between 0.448 and 0.5009, 
whereas the absorbed power scores 0.2381.
Hence the representation-learning model picks up on 
more of the spin glass's discriminatory ability
than the absorbed power does.

In summary, a many-body system's ability to discriminate amongst
combinations of familiar fields
can be quantified with the score of maximum-likelihood estimates
formed from a neural network's latent space.

\subsection*{Novelty detection: Has the system encountered this drive before?}
\label{sec_ROC}


\begin{figure}[hbt]
\centering
\includegraphics[width=.5\textwidth, clip=true]{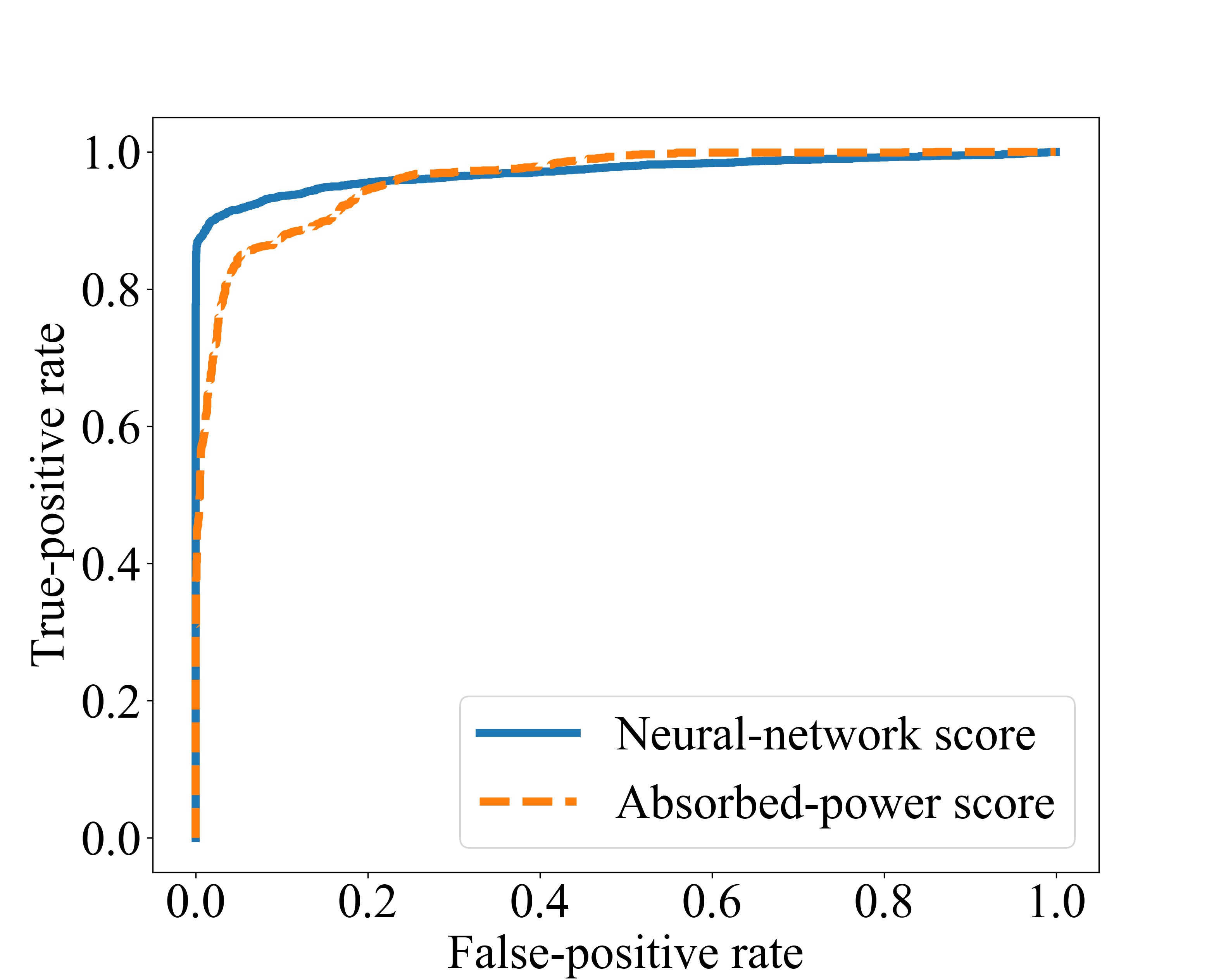}
\caption{\caphead{Receiver-operating-characteristic (ROC) curve:}
The spin glass was trained with three drives,
then tested with a familiar drive or with a novel drive.
From a response of the system's, an ROC curve can be defined.
The blue, solid curve is defined in terms of a bottleneck neural network;
and the orange, dashed curve is defined in terms of absorbed power.
}
\label{fig_ROC}
\end{figure}

At the start of the introduction, we described
how absorbed power has been used to identify novelty detection.
A system detects novelty when labeling a stimulus as familiar or unfamiliar.
The stimulus produces a response that exceeds a threshold or lies below.
If the stimulus exceeds the threshold, 
an observer should guess that the stimulus is novel.
Otherwise, the observer should guess that the stimulus is familiar.

The observer can err in two ways:
One commits a \emph{false positive} by believing a familiar drive to be novel.
One commits a \emph{false negative} by believing a novel drive to be familiar.
The errors trade off:
Raising the threshold lowers the probability
$p( \text{pos.} | \text{neg.} )$,
suppressing false positives at the cost of false negatives.
Lowering the threshold lowers the probability 
$p( \text{neg.} | \text{pos.} )$,
suppressing false negatives at the cost of false positives.

The \emph{receiver-operating-characteristic} (ROC) curve
depicts the tradeoff's steepness 
(see~\cite{Brown_06_Receiver} and Fig.~\ref{fig_ROC}).
Each point on the curve corresponds to one threshold value.
The false-positive rate $p( \text{pos.} | \text{neg.} )$ 
runs along-the $x$-axis; and the true-positive rate, 
$p( \text{pos.} | \text{pos.} )$, along the $y$-axis.
The greater the area under the ROC curve,
the more sensitively the response reflects accurate novelty detection.

We measure a many-body system's 
novelty-detection ability using an ROC curve.
Let us illustrate with the spin glass.
We constructed two random drives,
$\{A, B, C \}$ and $\{D, E, F \}$.
We trained the spin glass on $\{A, B, C \}$.
In each of 3,000 trials, we then tested the spin glass with
$A$, $B$, or $C$.
In each of 3,000 other trials, we tested with $D$, $E$, or $F$.
We defined one response in terms of a bottleneck neural network, as detailed below;
measured the absorbed power;
and, from each response, drew an ROC curve
(Fig.~\ref{fig_ROC}).
The curves show that representation learning and absorbed power 
reflect the spin glass's novelty detection about equally well. 
Each method excels slightly in one regime or another.

We defined the representation-learning response as follows.
We trained the neural network on the configurations 
assumed by the spin glass during its training.
The neural network populated latent space with three clumps of dots.
We modeled the clumps with a hard mixture 
$p_{ABC} (z_1,  z_2)$ of three Gaussians.\footnote{
(A mixture is hard if it models each point as
belonging to only one Gaussian.)}
We then fed the neural network the configuration 
that resulted from testing the spin glass.
The neural network mapped the configuration to a latent-space point 
$(z_1^{\rm test},  z_2^{\rm test})$.
We calculated the probability 
$p_{ABC} (z_1^{\rm test},  z_2^{\rm test}) \, dz_1 dz_2$ 
that the $ABC$ distribution produced the new point.
This probability was compared to a fixed threshold.
If the probability exceeded the threshold, 
the test configuration was guessed to have been produced by a novel drive.
We repeated this protocol with the other test trials, using the fixed threshold.
The fraction of guesses that were true positives,
and the fraction of guesses that were false positives,
specified one point on the blue, solid curve in Fig.~\ref{fig_ROC}.
Varying the threshold led to the other points.

We defined a thermodynamic ROC curve in terms of absorbed power.
Consider the trials in which the spin glass is tested with field $A$.
We histogrammed the power absorbed by the spin glass 
after the $A$ test.
We formed another histogram from the $B$-test trials;
and a third histogram, from the $C$-test trials.
To these histograms was compared
the power $\mathcal{P}$ that the spin glass absorbed
during a test with an arbitrary field.
We inferred the likelihood that $\mathcal{P}$ resulted from
a familiar field.
The results form the orange, dashed curve in Fig.~\ref{fig_ROC}.

The two ROC curves enclose regions of approximately the same area:
The neural network curve encloses an area-0.9633 region;
and the thermodynamic curve, an area-0.9601 region.
On average across all thresholds, therefore,
the responses register novelty detection approximately equally.
Yet the responses excel in different regimes:
The neural network achieves greater true-positive rates at low false-positive rates,
and the absorbed power achieves greater true-positive rates 
at high false-positive rates.
This two-regime behavior persisted across batches of trials,
though the enclosed areas fluctuated slightly.
Hence anyone paranoid about avoiding false positives
should measure a many-body system's novelty detection with a neural network.
Those more relaxed might prefer the absorbed power.

Why should the neural network excel at low false-positive rates?
Because of the neural network's skill at generalizing, we expect.
Upon training on cat pictures, a neural network generalizes from the instances.
Shown a new cat, the neural network recognizes its catness.
Perturbing the input a little perturbs the neural network's response little.
Hence changing the magnetic field a little,
which changes the spin-glass configuration little,
should change latent space little,
obscuring the many-body system's novelty detection.
This obscuring disappears when the magnetic field changes substantially.

In summary, a many-body system's novelty-detection ability is quantified with
an ROC curve formed from a neural network's latent space
or a thermodynamic response, depending on the false-positive threshold.

\subsection*{Feasibility}

Applying our toolkit might appear impractical,
since microstates must be inputted into the neural network.
Measuring a many-body system's microstate may daunt experimentalists.
Yet the use of microstates hinders our proposal little, for three reasons.

First, microstates can be calculated in numerical simulations,
which inform experiments.
Second, many key properties of many-body microstates 
have been measured experimentally.
High-speed imaging has been used to monitor soap bubbles' positions~\cite{Mukherji_19_Strength}
and colloidal suspensions~\cite{Cheng_11_Imaging}.
Similarly wielded tools, such as high magnification, have advanced
active-matter~\cite{Sanchez_12_Spontaneous}
and gene-expression~\cite{Lonsdale_13_Genotype} studies.

One might worry that the full microstate
cannot be measured accurately or precisely.
Soap bubbles' positions can be measured with finite precision,
and other microscopic properties might be inaccessible.
But, third, some bottleneck neural networks denoise their inputs~\cite{Vincent_08_Extracting,Goodfellow_16_Deep}:
The neural networks learn the distribution from which samples are generated ideally,
not systematic errors.
Denoising by variational autoencoders is less established but is progressing~\cite{Im_15_Denoising}.


%
%
\section*{Acknowledgements}
\noindent The authors thank Alexander Alemi, Isaac Chuang, Emine Kucukbenli, Nick Litombe, Seth Lloyd, Julia Steinberg, Tailin Wu, and Susanne Yelin for useful discussions.
WZ is supported by ARO Grant W911NF-18-1-0101; 
the Gordon and Betty Moore Foundation Grant, under No. GBMF4343;
and the Henry W. Kendall (1955) Fellowship Fund.
JMG is funded by the AFOSR, under Grant FA9950-17-1-0136.
SM was supported partially by the Moore Foundation, 
via the Physics of Living Systems Fellowship.
This material is based upon work supported by, or in part by, the Air Force
Office of Scientific Research, under award number FA9550-19-1-0411.
JLE has been funded by the Air Force Office of Scientific Research grant FA9550-17-1-0136 and by the James S. McDonnell Foundation Scholar Grant 220020476.
NYH is grateful for an NSF grant for the Institute for Theoretical Atomic, Molecular, and Optical Physics at Harvard University and the Smithsonian Astrophysical Observatory.
NYH also thanks CQIQC at the University of Toronto, the Fields Institute, and Caltech's Institute for Quantum Information and Matter (NSF Grant PHY-1733907) for their hospitality during the development of this paper.

%
%
%
\section*{Data availability and code availability}
\noindent 
The machine-learning code is available at~\cite{Github_repo}.
Spin-glass--simulation code will be available at~\cite{Github_repo}
once COVID-19 restrictions loosen enough that 
we can access the computers that store the files.

\begin{appendices}

\onecolumngrid

\renewcommand{\thesection}{\Alph{section}}
\renewcommand{\thesubsection}{\Alph{section} \arabic{subsection}}
\renewcommand{\thesubsubsection}{\Alph{section} \arabic{subsection} \roman{subsubsection}}

\makeatletter\@addtoreset{equation}{section}
\def\theequation{\thesection\arabic{equation}}

\section{Details about the bottleneck neural network}
\label{sec_NN_Details}


We briefly motivate and review variational autoencoders,
then describe the variational autoencoder applied in the main text.
Further background about variational autoencoders can be found in~\cite{Kingma_13_Auto,JR_14_Stochastic,Doersch_16_Tutorial}.
We denote vectors with boldface in this appendix.

Denote by $\mathbf{X}$ data that has a probability 
$p_{ \bm{\theta} }(\mathbf{x})$
of assuming the value $\mathbf{x}$.
$\bm{\theta}$ denotes a parameter,
and $p_{ \bm{\theta} } (\mathbf{x})$ is called the \emph{evidence}.
We do not know the form of $p_{ \bm{\theta} }(\mathbf{x})$,
when using representation learning.
We model $p_{ \bm{\theta} }(\mathbf{x})$ by identifying 
latent variables $\mathbf{Z}$ 
that assume the possible values $\mathbf{z}$.
Let $p_{ \bm{\theta} }(\mathbf{x} | \mathbf{z})$ 
denote the conditional probability that
$\mathbf{X} = \mathbf{x}$, given that $\mathbf{Z} = \mathbf{z}$.
We model the evidence, using the latent variables, with
\begin{align}
 p_{ \bm{\theta} }(\mathbf{x}) 
 = \int d\mathbf{z} \; 
 p_{ \bm{\theta} }( \mathbf{x} | \mathbf{z} ) 
 p( \mathbf{z} ) .
\end{align}

$p_{ \bm{\theta} }(\mathbf{x} | \mathbf{z} )$ 
can be related to the posterior distribution 
$p_{ \bm{\theta} } (\mathbf{z} | \mathbf{x})$.
The posterior is the probability that, if 
$\mathbf{X} = \mathbf{x}$, then $\mathbf{Z} = \mathbf{z}$. By Bayes' rule,
$p_{ \bm{\theta} }( \mathbf{z} | \mathbf{x} ) 
= p_{ \bm{\theta} }( \mathbf{x} | \mathbf{z} )
p( \mathbf{z} )  /  p_{ \bm{\theta} }( \mathbf{x} )$. 
Calculating the posterior is usually impractical,
as $p_{ \bm{\theta} }(\mathbf{x})$ 
is typically intractable (cannot be calculated analytically).
Hence we approximate the posterior with a variational model 
$q_{ \bm{\phi} }( \mathbf{z} | \mathbf{x} )$.
The optimization parameter $\bm{ \phi }$ denotes 
the neural network's weights and biases.

The approximation introduces an inference error,
quantified with the Kullback-Leibler divergence.
Let $P(\mathbf{u})$ and $Q(\mathbf{u})$ denote distributions over
the possible values $\mathbf{u}$ of a variable.
The Kullback-Leibler divergence quantifies 
the distance between the distributions:
\begin{align}
   D_\KL \LParen P(\mathbf{u}) || Q(\mathbf{u}) \RParen
   & :=  \mathbb{E}_{ P(\mathbf{u}) }
        \left[  \ln P( \mathbf{u} ) \right]
        -  \mathbb{E}_{ P(\mathbf{u}) }
        \left[ \ln  Q(\mathbf{u})  \right] \\
   \label{eq_D_KL_Nonneg}
   & \geq  0 .
\end{align}
$\mathbb{E}_{ P(\mathbf{u}) } [ f( \mathbf{u} ) ]$ denotes
the average of a function $f(\mathbf{u})$ over the distribution $P( \mathbf{u} )$.
Operationally, the Kullback-Leibler divergence equals the maximal efficiency 
with which the distributions can be distinguished, 
on average, in a binary hypothesis test.
We quantify our inference error with 
the Kullback-Leibler divergence between
the variational model and the posterior,
$D_\KL \LParen  
q_{ \bm{\phi} }( \mathbf{z} | \mathbf{x} ) || 
                         p_{ \bm{\theta} }( \mathbf{z} | \mathbf{x} )  \RParen$. 

Recall that we wish to estimate $p_{\bm \theta} ( \mathbf{x} )$:
An accurate estimate lets us predict $\mathbf{x}$ accurately.
We wish also to estimate the latent posterior distribution, 
$q_{\bm \phi} (\mathbf{z} | \mathbf{x} )$.
We therefore write out the Kullback-Leibler divergence's form,
apply Bayes' rule to rewrite the $p_{ \bm{\theta} }( \mathbf{z} | \mathbf{x} )$,
rearrange terms,
and repackage terms into a new Kullback-Leibler divergence:
\begin{align}
   \label{eq_Log_Like_1}
   \ln p_{ \bm{\theta} } ( \mathbf{x} )
   =  D_\KL \LParen  q_{\bm \phi} ( \mathbf{z} | \mathbf{x} )
                                                     || p_{\bm \theta} ( \mathbf{z} | \mathbf{x} )
                   \RParen
   +  \mathbb{E}_{ q_{\bm \phi} ( \mathbf{z} | \mathbf{x} )}
   \left[ \ln p_{\bm \theta} ( \mathbf{x} | \mathbf{z} )  \right]
   -  D_\KL  \LParen  q_{\bm \phi} ( \mathbf{z} | \mathbf{x} )
                                                       || p ( \mathbf{z} )  \RParen .
\end{align}
The penultimate term encodes our first goal;
and the final term, our second goal.

Recall that the Kullback-Leibler divergence is nonnegative.
The sum of the final two terms therefore lower-bounds the log-likelihood,
$\ln p_{\bm \theta}( \mathbf{x} )$.
$\mathbf{x}$ denotes the event observed,
$\bm{\theta}$ denotes a possible cause,
and $p_{\bm \theta}$ denotes the likelihood that 
$\bm{\theta}$ caused $\mathbf{x}$.
Maximizing each side of Eq.~\eqref{eq_Log_Like_1},
and invoking Ineq.~\eqref{eq_D_KL_Nonneg}, yields
\begin{align}
   \label{eq_Log_Like_2}
   \max_\theta  \left\{   \ln p_{ \bm{\theta} } ( \mathbf{x} )   \right\}
   \geq \max_\theta  \left\{
   \mathbb{E}_{ q_{\bm \phi} ( \mathbf{z} | \mathbf{x} )}
   \left[ \ln p_{\bm \theta} ( \mathbf{x} | \mathbf{z} )  \right]
   -  D_\KL  \LParen  q_{\bm \phi} ( \mathbf{z} | \mathbf{x} )
                                                       || p ( \mathbf{z} )  \RParen 
   \right\} .
\end{align}
The RHS is called the \textit{evidence lower bound} (ELBO). 

A variational autoencoder is a neural network that implements the ELBO.
$q_{\bm \phi} ( \mathbf{z} | \mathbf{x} )$ encodes the input $\mathbf{X}$, 
and $p_{\bm \theta} ( \mathbf{x} | \mathbf{z} )$ decodes.
The variational autoencoder has the cost function
\begin{equation}
   \label{eq_VAE_Cost}
   \mathcal{L}_{\text{VAE}} 
   := \mathbb{E}_{  p_{\rm emp}( \mathbf{x} )  }
   \left[ \mathbb{E}_{  q_{\bm \phi} ( \mathbf{z} | \mathbf{x} )  }
   \left[  \ln p_{\bm \theta} ( \mathbf{x} | \mathbf{z} )  \right] 
   -  D_\KL  \LParen  q_{\bm \phi} ( \mathbf{z} | \mathbf{x} ) 
                                                        \| p( \mathbf{z} )  \RParen \right].
\end{equation}
$p_{\rm emp}( \mathbf{x} )$ denotes the distribution inferred from the empirical dataset.
Given input values $\mathbf{x}$, 
the variational autoencoder generates a latent distribution 
$q_{\bm \phi} ( \mathbf{z} | \mathbf{x} ) 
= \mathcal{N}( \bm{\mu}_{ \mathbf{z} | \mathbf{x} }, 
                                           \bm{\Sigma}_{ \mathbf{z} | \mathbf{x} } )$.
We denote by $\mathcal{N} ( \bm{\mu}, \bm{\Sigma} )$
the standard multivariate normal distribution
whose vector of means is $\bm{\mu}$ 
and whose covariance matrix is $\bm{\Sigma}$.
Neural-network layers parameterize the variational autoencoder's
$\bm{\mu}_{ \mathbf{z} | \mathbf{x} }$ and 
$\bm{\Sigma}_{ \mathbf{z} | \mathbf{x} }$.
Latent vectors are sampled according to 
$q_{\bm \phi}( \mathbf{z} | \mathbf{x} )$,
then decoded into outputs distributed according to
$p_{\bm \theta} ( \mathbf{x} | \mathbf{z} ) 
= \mathcal{N}( \bm{\mu}_{ \mathbf{x} | \mathbf{z} }, 
                        \sigma^2_{ \mathbf{x} | \mathbf{z} }  \id  )$.
Neural-network layers parameterize the mean vector 
$\bm{\mu}_{ \mathbf{x} | \mathbf{z} }$.
The variance $\sigma^2_{ \mathbf{x} | \mathbf{z} }$ is a hyperparameter.

A variational autoencoder with the following architecture 
produced the results in the main text.
The style was borrowed from~\cite{Hafner_18_Building}.
Two fully connected $200$-neuron hidden layers
process the input data.
One fully connected two-neuron hidden layer parameterizes each of
$\bm{\mu}_{ \mathbf{z} | \mathbf{x} }$ and 
$\bm{\Sigma}_{ \mathbf{z} | \mathbf{x} }$.
Two fully connected $200$-neuron hidden layers process the latent variables.
An output layer reads off the outputs. 
We choose $\sigma^2_{ \mathbf{x} | \mathbf{z} }=1$ 
and use Rectified Linear Unit (ReLU) activations
for all hidden layers.

\section{Justification of use of machine learning}
\label{sec_Justify_ML}


Deep learning is a powerful tool.
Is it necessary for recovering our results?
Could simpler algorithms detect and quantify many-body learning as sensitively?
Comparable simpler algorithms tend not to, we find.
Two competitors suggest themselves:
single-layer linear autoencoders,
related to principal-component analysis~\cite{Bourland_88_Auto}, 
and clustering algorithms.
Alternatives include generalized linear models~\cite{Bishop_06_Pattern}
and supervised linear autoencoders.
These models, however, perform supervised learning.
They receive more information than the variational autoencoder
and so enjoy an unfair advantage.
Furthermore, supervised learners receive information not directly available 
to the many-body system---the drives' labels (see Results).
One therefore cannot infer, from a supervised learner's output,
how much the many-body system has learned.
We analyze the two comparable competitors sequentially.\footnote{
Alternative representations of inputs have been defined
to facilitate predictions, e.g.,~\cite{Shalizi_01_Computational}.
We seek a representation (e.g., a latent space) that elucidates how the spin glass's configurations reflect the drive. 
This problem differs from that addressed in~\cite{Shalizi_01_Computational}.
Nevertheless, modifications of recent techniques of Crutchfield \emph{et al.} may provide useful alternatives to the standard machine-learning toolkit in future work.}

%
%
%
\subsection{Comparison with single-layer linear autoencoder}

The linear autoencoder is a single-layer neural network.
The input, $X$, undergoes a linear transformation:
$Y = mX + b$.
We compare, as follows, the linear autoencoder's 
detection of field classification 
with the variational autoencoder's detection:
We trained the spin glass on a drive in each of 
3,000-5,000 trials.
Ninety percent of the trials were designated as neural-network-training data;
and 10\%, as neural-network-testing data.
For each training trial, we identified the spin glass's final configuration.
On these configurations, 
each neural network performed unsupervised learning.
Each neural network then received the configuration with which
the spin glass ended a neural-network-testing trial.
We inferred the field most likely to have produced this configuration,
using maximum-likelihood estimation.
The fraction of trials in which the neural network points to the correct field 
constitutes the neural network's score.
On a three-field drive, the linear autoencoder scored 0.771,
while the variational autoencoder scored 0.992.
On a five-field drive, the linear autoencoder scored 0.3934,
while the variational autoencoder scored 0.829.
Hence the variational autoencoder picks up on more of 
the spin glass's ability to classify fields.

\subsection{Comparison with clustering algorithm}

A popular, straightforward-to-apply algorithm is 
\emph{$k$-means clustering}~\cite{Bishop_06_Pattern}.
$k$ refers to a parameter inputted into the algorithm,
the number of clusters expected in the data.
We inputted the number of drives imposed on the spin glass,
in addition to inputting configurations.
The variational autoencoder receives just configurations 
and so less information.
We could level the playing field by automating the choice of $k$,
using the Bayesian information criterion (BIC)~\cite{Bishop_06_Pattern}.
But clustering with the BIC-chosen $k$ 
would perform no better than
clustering performed with the ideal $k$,
and the ideal clustering performs worse than the variational autoencoder.

The protocol run on the spin glass is described 
at the beginning of Methods,
in the memory-capacity study.
Five thousand trials were performed.
The configuration occupied by the spin glass 
at the end of each trial was collected.
Splitting the data into testing and training data
did not alter results significantly.
Hence we fed all the configurations,
with the number $k = 5$ of drives,
to the clustering algorithm.
The algorithm partitioned the set of configurations into subsets.
Each subset contained configurations 
likely to have resulted from the same drive.

Clustering algorithms are assessed with the Rand index,
denoted by RI~\cite{Rand_71_Objective}.
The Rand index differs from the maximum-likelihood-estimation score 
(discussed in the Results).
How to compare the clustering algorithm
with the variational autoencoder, therefore, is ambiguous.
However, the Rand index quantifies 
the percentage of the algorithm's classifications that are correct.
Hence the Rand index and the maximum-likelihood-estimation score
have similar interpretations, despite their different definitions.

The clustering algorithm's Rand index began at $\text{RI} = 0$, at $t = 0$.
RI rose during the first $\approx 200$ changes of the drive,
then oscillated around 0.125.
Figure~\ref{fig_Capacity} shows the variational autoencoder's performance.
The variational autoencoder's score rose during 
the first $\approx 150$ changes of the drive,
then oscillated around $0.450 > 0.125$.
Hence the variational autoencoder outperformed the clustering algorithm.

\section{Distinction between robust learning \\
and two superficially similar behaviors}
\label{sec_Not_Enslaved_Or_Frozen}


Learning contrasts with two other behaviors
that the spin glass could exhibit,
entraining to the field and near-freezing.

\subsection{Entraining to the field}
\label{sec_Enslave}

Imagine that most spins align with any field $A$. 
The configuration reflects the field
as silly putty reflects the print of a thumb
pressing on the silly putty.
Smoothing the silly putty's surface wipes the thumbprint off.
Similarly, applying a field $B \neq A$ to the spin glass
wipes the signature of $A$ from the configuration.
From the perspective of the end of the application of $B$,
the spin glass has not learned $A$.
The spin glass lacks a long-term memory of the field;
the field is encoded in no robust, deep properties of the configuration.

We can distinguish learning from entraining
by calculating the percentage of the spins 
that align with the field at the end of training.
If the spins obeyed the field, 100\% would align.
If the spins ignored the field, $50\%$ would align, on average.
Hence the spin glass's entraining is quantified with
\begin{align}
   2 ( \text{percentage of spins aligned with the field} )  - 100 .
\end{align}
(This measure does not apply to alignment percentages $< 50$,
which are unlikely to be realized.)

\begin{figure}[hbt]
\centering
\includegraphics[width=.45\textwidth, clip=true]{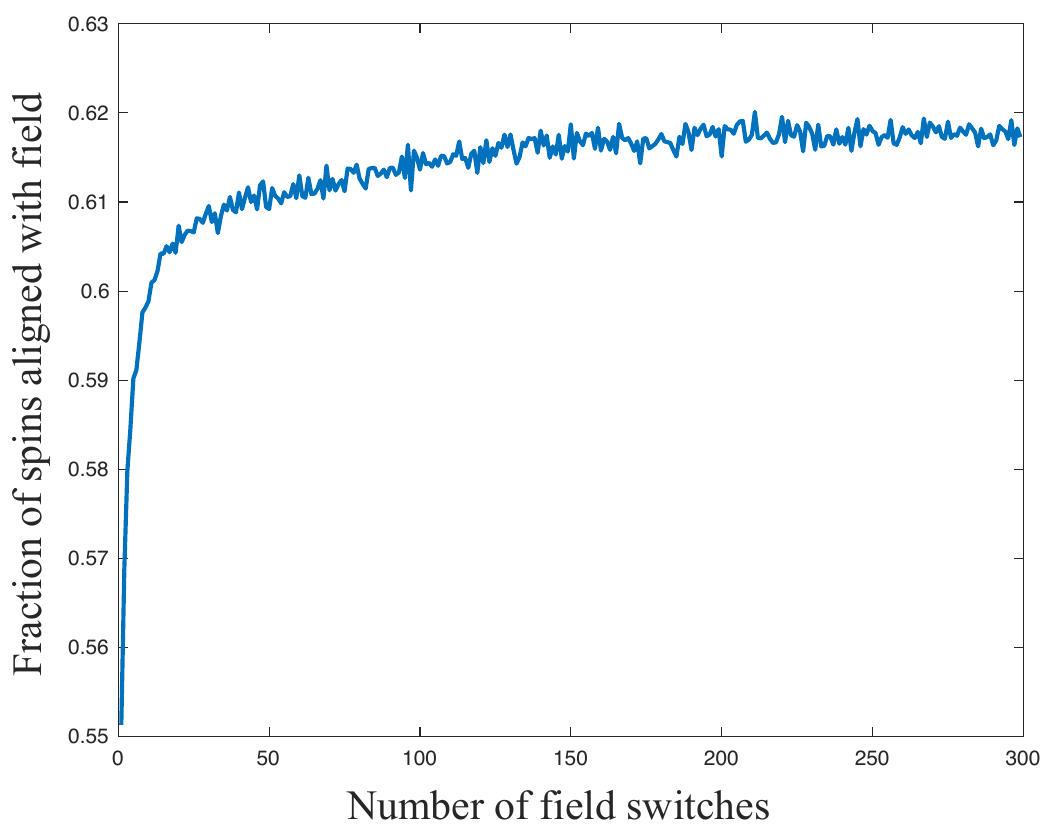}
\caption{\caphead{Fraction of the spins aligned with the field,
as a function of time:}
If a fraction $\approx 1$ of the spins align, the spin glass resembles silly putty,
which shallowly reflects the print of a thumb that presses on it.
Robust learning stores information deep in a system's structure.}
\label{fig_Frac_Aligned}
\end{figure}

Figure~\ref{fig_Frac_Aligned} shows data collected 
about the spin glass in the good-learning regime 
(introduced in the ``Spin glass'' section of the Results).
The number of aligned spins is plotted against
the amount $t$ of time for which the spin glass has trained.
After the application of one field, 55\% of the spins align with the field.
At the end of training, 62\% align. 
Hence the spins' entraining grows from 10\% to 24\%.
Growth is expected, as the spin glass learns the training drive.
But $24\%$ is an order of magnitude less than $100\%$, 
so the spin glass is not entrained to the field.

\subsection{Near-freezing}
\label{sec_Near_Freeze}

Suppose that the spin glass is nearly frozen.
Most spins cannot flip, but a few jiggle under most fields.
The spin glass does not learn any field effectively,
being mostly immobile.
But the few flippable spins reflect the field.
A bottleneck neural network could guess the field from those few spins.
The neural network's low loss function would induce a false positive, 
leading us to believe that the spin glass had learned.


We can avoid false positives by measuring two properties.
First, we measure the percentage of the spins
that antialign with the field.
If the percentage consistently $\gg 0$,
many of the spins are not frozen.
Figure~\ref{fig_Frac_Aligned} confirms that many are not.

\begin{figure}[hbt]
\centering
\includegraphics[width=.5\textwidth, clip=true]{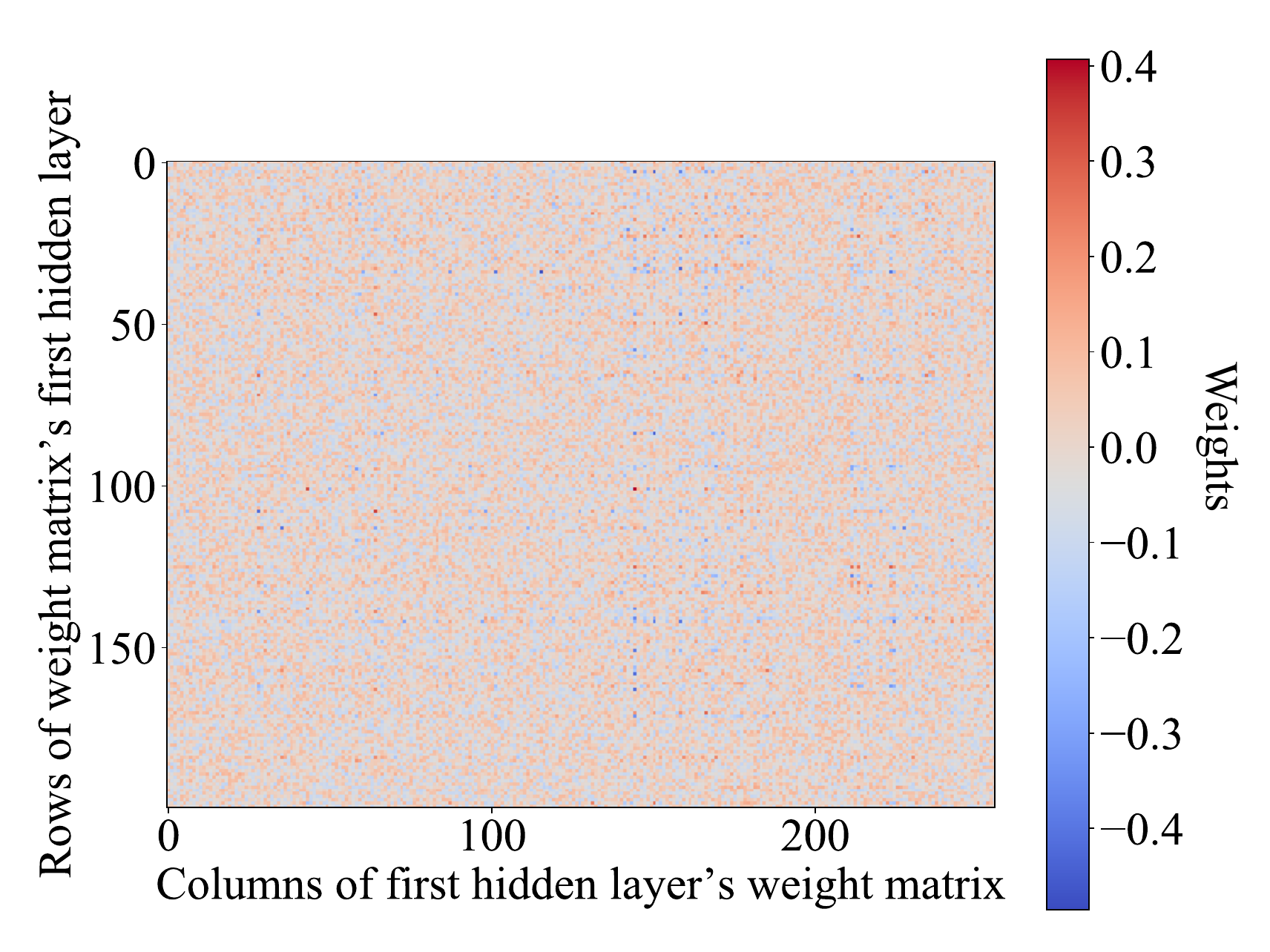}
\caption{\caphead{How much information about each spin 
the variational autoencoder compresses:}
This figure represents the first hidden layer's weight matrix.
The weight matrix transforms the input layer, which consists of 256 neurons,
into the first hidden layer, which consists of 200 neurons.
The matrix's elements are replaced with colors.
Each vertical line corresponds to one spin.
The farther leftward a stripe, the lesser the spin's field energy 
[Eq.~\eqref{eq_Hamiltonian_j}].
} 
\label{fig_Which_Spins_Used_By_AE_b}
\end{figure}

Second, we check that the neural network compresses information about
spins that have many different field energies $A_j(t) s_j$
[Eq.~\eqref{eq_Hamiltonian_j}].
We illustrate with the protocol used to generate Fig.~\ref{fig_Latent_Space}:
We trained the spin glass on a drive $\{A, B, C\}$
in each of many trials.
On the end-of-trial configurations, the neural network was trained.

A configuration is represented in the neural network's input layer, 
a column vector.
A weight matrix transforms the input layer
into the first hidden layer, another column vector.
The weight matrix is depicted in Fig.~\ref{fig_Which_Spins_Used_By_AE_b}.
The matrix's numerical entries have been replaced with colors.
Each vertical stripe corresponds to one spin.
The farther leftward a stripe, the lesser the spin's field energy.
The darker a stripe, the more information about the spin 
the neural network uses when forming $Z$.
The plot is approximately invariant, at a coarse-grained level,
under translations along the horizontal.
(On the order of ten exceptions exist.
These vertical stripes contain several dark dots.
An example appears at $x \approx 150$.
But the number of exceptions is much less than the number of spins:
$\approx 10 \ll 256$.)
Hence the neural network uses information about spins of many field energies.
The spins do not separate into
low-field-energy flippable spins
and high-field-energy frozen spins.

\section{Memory capacity attributed to the many-body system \\
by the absorbed power}
\label{app_Capacity_Work}


In the Methods, we compared the memory capacity
registered by the neural network
to the capacity registered by the absorbed power.
The study involved maximum-likelihood estimation on 
drives of 40 fields selected from 50 fields.
The choice of 50 is explained here:
Fifty fields exceed the spin-glass capacity
registered by the absorbed power.

Recall how memory has been detected thermodynamically~\cite{Gold_19_Self}:
Let a many-body system be trained with
a drive that includes a field $A$.
Consider testing the system, afterward, 
with an unfamiliar field $B$, and then with $A$.
Suppose that the absorbed power jumps substantially when $B$ is applied
and less when $A$ is reapplied.
The many-body system identifies $B$ as novel
and remembers $A$,
according to the absorbed power.

We sharpen this analysis.
First, we divide the trial into time windows.
During each time window, the field switches 10 times.
(The 10 eliminates artificial noise and is not critical.
Our qualitative results are robust with respect to changes in such details.)
We measure the absorbed power at the end of each time window
and at the start of the subsequent window.
We define ``the absorbed power jumps substantially'' as
``the absorbed power jumps, on average over trials, 
by much more than the noise
(by much more than the absorbed power fluctuates across a trial)'':
\begin{align}
   \label{eq_Capacity_Condn_Work}
   & \langle ( \text{Power absorbed at start of later time window} ) 
   - ( \text{Power absorbed at end of preceding time window} )
   \rangle_{ \text{trials} } 
   \\ \nonumber & 
   \gg  \text{Standard deviation in }
   [ ( \text{Power absorbed at start of later window} )
   -  ( \text{Power absorbed at end of preceding window} ) ] .
\end{align}
Consider including only a few fields in the training drive, 
then growing the drive in later trials.
The drive will tax the spin glass's memory until exceeding the capacity.
The LHS of~\eqref{eq_Capacity_Condn_Work}
will come to about equal the RHS.

\begin{figure}[hbt]
\centering
\includegraphics[width=.45\textwidth, clip=true]{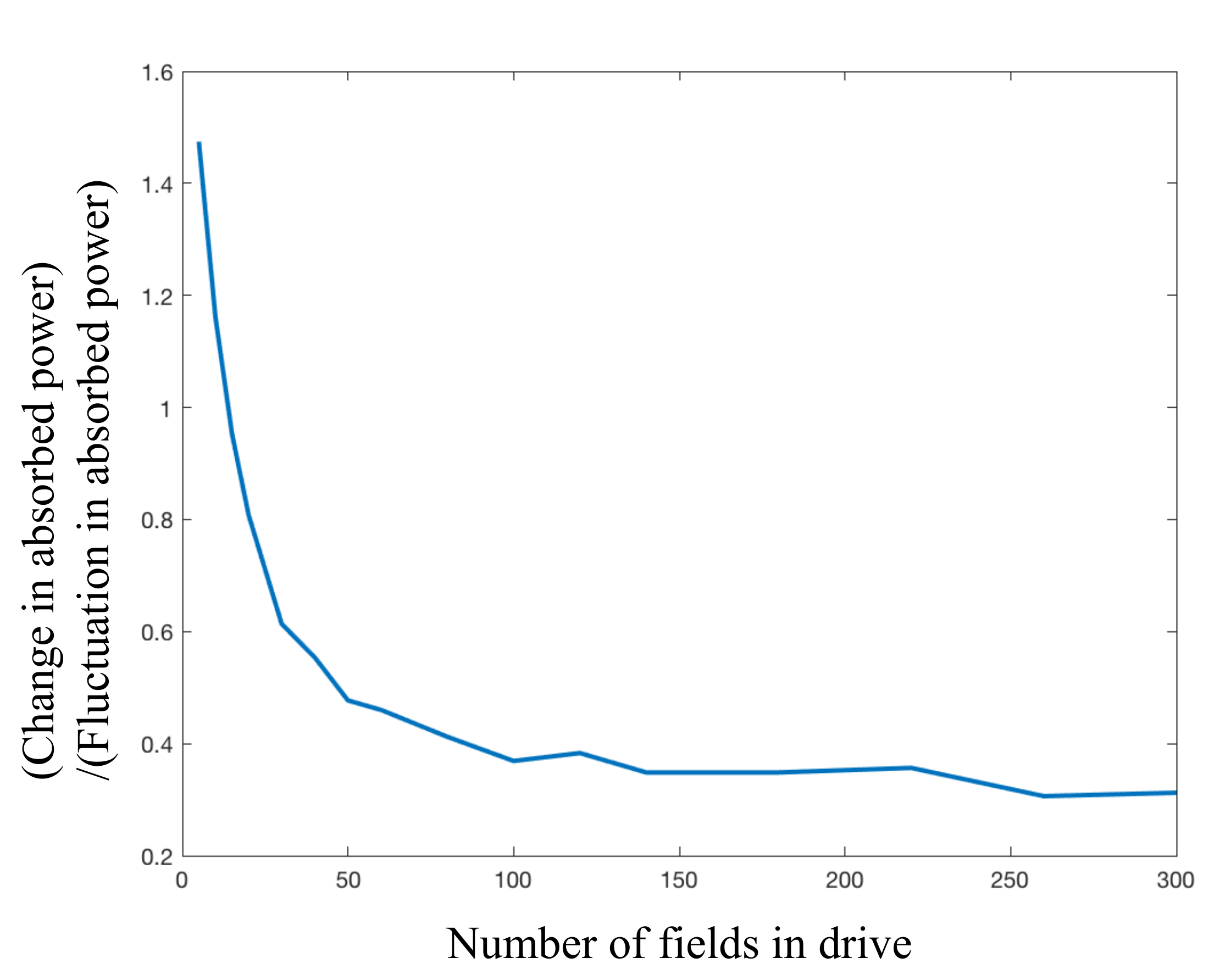}
\caption{\caphead{Estimate of memory capacity by absorbed power:}
A many-body system reaches its capacity, 
according to the absorbed power, when
[left-hand side of Ineq.~\eqref{eq_Capacity_Condn_Work}]
/ (right-hand side) $\approx 1$.
The curve $\approx 1$, and a 256-spin glass reaches its capacity,
when the training drive contains $\approx 15$ fields.}
\label{fig_Capacity_Work}
\end{figure}

Figure~\ref{fig_Capacity_Work} illustrates with the spin glass.
On the $x$-axis is the number of fields in the training drive.
On the $y$-axis is the ratio of 
the left-hand side of Ineq.~\eqref{eq_Capacity_Condn_Work}
to the right-hand side (LHS/RHS).
Where LHS/RHS $\approx 1$, the spin glass reaches its capacity.
This spin glass can remember $\approx 15$ fields,
according to the absorbed power.

\end{appendices}

%
%
\bibliographystyle{h-physrev}
\bibliography{Noneq_ML_Bib}

\begin{thebibliography}{10}

\bibitem{Coppersmith_97_Self}
S.~N. Coppersmith {\em et~al.},
\newblock Phys. Rev. Lett. {\bf 78}, 3983 (1997).

\bibitem{Povinelli_99_Noise}
M.~L. Povinelli, S.~N. Coppersmith, L.~P. Kadanoff, S.~R. Nagel, and S.~C.
  Venkataramani,
\newblock Phys. Rev. E {\bf 59}, 4970 (1999).

\bibitem{Keim_11_Generic}
N.~C. Keim and S.~R. Nagel,
\newblock Phys. Rev. Lett. {\bf 107}, 010603 (2011).

\bibitem{Keim_13_Multiple}
N.~C. Keim, J.~D. Paulsen, and S.~R. Nagel,
\newblock Phys. Rev. E {\bf 88}, 032306 (2013).

\bibitem{Paulsen_14_Multiple}
J.~D. Paulsen, N.~C. Keim, and S.~R. Nagel,
\newblock Phys. Rev. Lett. {\bf 113}, 068301 (2014).

\bibitem{Majumdar_18_Mechanical}
S.~Majumdar, L.~C. Foucard, A.~J. Levine, and M.~L. Gardel,
\newblock Soft Matter {\bf 14}, 2052 (2018).

\bibitem{Mukherji_19_Strength}
S.~Mukherji, N.~Kandula, A.~K. Sood, and R.~Ganapathy,
\newblock Phys. Rev. Lett. {\bf 122}, 158001 (2019).

\bibitem{Zhong_17_Associative}
W.~Zhong, D.~J. Schwab, and A.~Murugan,
\newblock Journal of Statistical Physics {\bf 167}, 806 (2017).

\bibitem{Keim_19_Memory}
N.~C. Keim, J.~D. Paulsen, Z.~Zeravcic, S.~Sastry, and S.~R. Nagel,
\newblock Rev. Mod. Phys. {\bf 91}, 035002 (2019).

\bibitem{Gold_19_Self}
J.~M. Gold and J.~L. England,
\newblock arXiv:1911.07216  (2019).

\bibitem{Nielsen_15_Neural}
M.~Nielsen,
\newblock {\em Neural Networks and Deep Learning} (Determination Press, 2015).

\bibitem{Goodfellow_16_Deep}
I.~Goodfellow, Y.~Bengio, and A.~Courville,
\newblock {\em Deep Learning} (MIT Press, 2016),
\newblock \url{http://www.deeplearningbook.org}.

\bibitem{Bengio_12_Representation}
Y.~{Bengio}, A.~{Courville}, and P.~{Vincent},
\newblock arXiv:arXiv:1206.5538  (2012).

\bibitem{Alemi_18_TherML}
A.~A. {Alemi} and I.~{Fischer},
\newblock arXiv:1807.04162  (2018).

\bibitem{Kingma_13_Auto}
D.~P. {Kingma} and M.~{Welling},
\newblock arXiv:1312.6114  (2013).

\bibitem{JR_14_Stochastic}
D.~{Jimenez Rezende}, S.~{Mohamed}, and D.~{Wierstra},
\newblock {Stochastic Backpropagation and Approximate Inference in Deep
  Generative Models},
\newblock in {\em Proc. 31st Int. Conf. on Machine Learning}, 2014.

\bibitem{Doersch_16_Tutorial}
C.~{Doersch},
\newblock arXiv:1606.05908  (2016).

\bibitem{Bourland_88_Auto}
H.~Bourlard and Y.~Kamp,
\newblock Biological Cybernetics {\bf 59}, 291 (1988).

\bibitem{Jaeger_02_Short}
H.~Jaeger,
\newblock Short-term memory in echo state networks, 2002.

\bibitem{Github_repo}
Online code,
\newblock https://github.com/smarzen/Statistical-Physics, 2020.

\bibitem{Bishop_06_Pattern}
C.~M. Bishop,
\newblock {\em Pattern Recognition and Machine Learning} (Springer, 2006).

\bibitem{Miller_19_Raft}
J.~Miller,
\newblock Physics Today  (2019).

\bibitem{Heylighen_02_Science}
F.~Heylighen,
\newblock {\em The Science of Self-Organization and Adaptivity} (EOLSS
  Publishers Co Ltd, 2002), chap.~7, pp. 184--211,
\newblock in: L. D. Kiel, (ed.) Knowledge Management, Organizational
  Intelligence and Learning, and Complexity, in: The Encyclopedia of Life
  Support Systems ((EOLSS), (Eolss Publishers, Oxford). [http://www.eolss.net]
  Series editor: L. D. Kiel, (ed.).

\bibitem{Carleo_19_Machine}
G.~Carleo {\em et~al.},
\newblock Rev. Mod. Phys. {\bf 91}, 045002 (2019).

\bibitem{Wu_19_Toward}
T.~Wu and M.~Tegmark,
\newblock Phys. Rev. E {\bf 100}, 033311 (2019).

\bibitem{Iten_20_Discovering}
R.~Iten, T.~Metger, H.~Wilming, L.~del Rio, and R.~Renner,
\newblock Phys. Rev. Lett. {\bf 124}, 010508 (2020).

\bibitem{Brown_06_Receiver}
C.~D. Brown and H.~T. Davis,
\newblock Chemometrics and Intelligent Laboratory Systems {\bf 80}, 24  (2006).

\bibitem{Cheng_11_Imaging}
X.~Cheng, J.~H. McCoy, J.~N. Israelachvili, and I.~Cohen,
\newblock Science {\bf 333}, 1276 (2011),
  https://science.sciencemag.org/content/333/6047/1276.full.pdf.

\bibitem{Sanchez_12_Spontaneous}
T.~Sanchez, D.~T.~N. Chen, S.~J. DeCamp, M.~Heymann, and Z.~Dogic,
\newblock Nature {\bf 491}, 431 (2012).

\bibitem{Lonsdale_13_Genotype}
J.~Lonsdale {\em et~al.},
\newblock Nature Genetics {\bf 45}, 580 (2013).

\bibitem{Vincent_08_Extracting}
P.~Vincent, H.~Larochelle, Y.~Bengio, and P.-A. Manzagol,
\newblock Extracting and composing robust features with denoising autoencoders,
\newblock in {\em Proceedings of the 25th International Conference on Machine
  Learning}, ICML '08, pp. 1096--1103, New York, NY, USA, 2008, ACM.

\bibitem{Im_15_Denoising}
D.~J. Im, S.~Ahn, R.~Memisevic, and Y.~Bengio,
\newblock CoRR {\bf abs/1511.06406} (2015).

\bibitem{Hafner_18_Building}
D.~Hafner,
\newblock Building variational auto-encoders in tensorflow,
\newblock Blog post, 2018.

\bibitem{Shalizi_01_Computational}
C.~R. Shalizi and J.~P. Crutchfield,
\newblock Journal of Statistical Physics {\bf 104}, 817 (2001).

\bibitem{Rand_71_Objective}
W.~M. Rand,
\newblock Journal of the American Statistical Association {\bf 66}, 846 (1971),
  https://www.tandfonline.com/doi/pdf/10.1080/01621459.1971.10482356.

\end{thebibliography}

\end{document}